\documentclass[preprint,12pt]{elsarticle}

\usepackage[a4paper,margin=1in]{geometry}
\usepackage{amsmath,amssymb,amsfonts,bm}
\usepackage{graphicx}
\usepackage{booktabs}
\usepackage{array}
\usepackage{xcolor}
\usepackage{pdfpages}
\usepackage{hyperref}
\usepackage{microtype}

\hypersetup{colorlinks=true,linkcolor=blue!70!black,citecolor=blue!70!black,urlcolor=blue!70!black}

\newcommand{\WriteInstantMaxDev}{0.006}
\newcommand{\WriteKeptLow}{0.65}
\newcommand{\WriteKeptHigh}{0.92}
\newcommand{\SweepSimRange}{0.839\text{--}0.843}
\newcommand{\SweepTheory}{0.841}

\newcommand{\LockLawActHoldFold}{138}
\newcommand{\LockLawActSpread}{23}

\newcommand{\SweptKramersUnderRange}{3\text{--}7}

\newcommand{\NetTotal}{180}

\newcommand{\NetPerKind}{60}

\newcommand{\PhaseSepEnd}{1.55}
\newcommand{\PhasePeriods}{39}

\newcommand{\PhaseLockSlope}{0.95}

\newcommand{\LoopCount}{91}

\newcommand{\SchloglFoldLow}{1.49}
\newcommand{\SchloglFoldHigh}{2.26}
\newcommand{\SchloglCuspA}{4.15}
\newcommand{\SchloglCuspB}{2.65}

\newcommand{\ToggleAlphaC}{2.00}

\newcommand{\ToggleOwnGamma}{0.31}
\newcommand{\ToggleLawMaxDev}{0.03}

\newcommand{\TubesMuC}{1.383}
\newcommand{\TubesUnequalMu}{1.21}

\newcommand{\RingThreeAlpha}{2.00}

\newcommand{\RingFourAlpha}{1.013}
\newcommand{\DipSRange}{0.04\text{--}0.05}

\newcommand{\ColloidStrongRewriteRange}{1.0}
\newcommand{\ColloidStrongLambdaMax}{0.75}
\newcommand{\ColloidWorkPerField}{5.0}

\newcommand{\OrderStates}{four}
\newcommand{\OrderClassesSplit}{three}

\newcommand{\RpmL}{32}
\newcommand{\RpmN}{1024}
\newcommand{\RpmR}{2}
\newcommand{\RpmSubloops}{144}
\newcommand{\RpmSamples}{24}

\newcommand{\RpmChainL}{512}

\newcommand{\BoundLandscapes}{2{,}305}
\newcommand{\BoundMaxRatio}{0.99}
\newcommand{\BoundWeakWithin}{8}

\newcommand{\BoundMobilityExcess}{7.7}
\newcommand{\BoundTemperatureExcess}{5.7}


\makeatletter\def\ps@pprintTitle{\let\@oddhead\@empty\let\@evenhead\@empty\let\@oddfoot\@empty\let\@evenfoot\@oddfoot}\makeatother

\begin{document}

\begin{frontmatter}

\title{Principles of Material Memory}

\author{Vladimir A. Baulin\corref{cor1}}
\ead{vbaulin@synthetix.institute}
\cortext[cor1]{Corresponding author}
\address{Synthetix Institute, Keizersgracht 241, 1016 EA Amsterdam, The Netherlands}

\begin{abstract}
Materials keep traces of how they were driven: a field leaves a magnetization behind, and a sequence of forces can leave a pattern that records the order in which they acted. It remains open what such a history records, which interactions create the record and which dynamics preserve it. This review relates the three. Forces applied in different orders can leave different states after the drive is removed, and in a spatial network the preferred local arrangements may not be reconcilable around a closed loop. Both effects compare two paths, one in time and one around a loop in space, and both are described by commutators, curvature and holonomy. Restoring forces and fluctuations then set how long the trace survives: a displacement relaxes, a phase diffuses, and barriers protect a pattern against rare thermal transitions. Near an instability, a weak writing bias is amplified while the coordinate that carries the state has no restoring force, and the barriers that retain the chosen state form only afterwards. The choice is therefore as accurate as the detection of the bias against the noise during the time for which the coordinate is free. Once a state is stored behind a barrier, a static bias replaces it through the same thermal channel by which it would be lost. The ratio of the mean retention time to the mean rewriting time is then bounded, up to a prefactor close to one, by the Boltzmann factor of an effective energy bias, whatever the height of the barrier. Calculations for capillary rotors and point-dipole arrays show how the geometry of the interactions constrains the patterns that a material can store. In a model of three coupled amplitudes, a weak bias makes two mirror-image patterns lose stability at slightly different coupling strengths, so that a coupling pulse between the two thresholds erases one pattern and preserves the other.
\end{abstract}

\begin{keyword}
Memory in matter \sep Metastability \sep Bifurcation \sep Geometric frustration \sep Holonomy \sep Stochastic switching
\end{keyword}

\end{frontmatter}

\tableofcontents

\section{Introduction}
\label{sec:introduction}

Ferromagnets return to states that record earlier reversals of the field~\cite{Keim2019Memory,PaulsenKeim2025}, because domain walls are pinned on quenched disorder~\cite{Barkhausen1919,Preisach1935,Sethna1993Hysteresis}. Structural glasses and polymer networks age, rejuvenate and recover asymmetrically after a change of temperature~\cite{Kovacs1979,Bouchaud1992Aging,Cugliandolo1993Aging}. Sheared non-Brownian suspensions settle into reversible orbits that record the amplitude of past shear~\cite{Corte2008Reversible,Keim2011Memory,Paulsen2014Memory}, and crumpled sheets record past compaction~\cite{Matan2002Crumpling}. Hydrogels with dynamic crosslinks store thermally written patterns as turbidity and lose them with tunable kinetics~\cite{Yu2020HydrogelMemory,Zhao2025GelHierarchy}. Mechanical metamaterials store inputs in bistable beams and folds~\cite{Silverberg2014Origami,Florijn2014Metamaterials,Coulais2016Metamaterials,Chen2021Reprogrammable}, and memristive materials store the history of the charge that has passed through them~\cite{Chua1971Memristor,Strukov2008Memristor,Yang2013Memristor}.

Many of these systems store the extreme values of a scalar drive, such as the largest strain or the largest magnetic field applied. The record shows itself when a later drive exceeds the earlier maximum and the response changes~\cite{Keim2019Memory,Sethna1993Hysteresis,Mayergoyz1991,Paulsen2014Memory}.

Memory is often attributed to the material substance, or to its governing differential equation alone. Yet the same equation of motion can retain a written state for many relaxation times or lose it within one, depending on a single control parameter and on the intensity of thermal noise. A stored distinction can also reside in internal coordinates to which a standard measurement is blind, so whether a material appears to remember depends on what is measured. A material can also record the temporal sequence in which several operations acted. Such a record requires operations that do not commute, but non-commutativity alone is not enough: a device that simply overwrites its state executes non-commuting operations and retains only the final input. These phenomena, observed in very different substrates, raise the same physical questions: how a distinction is written into matter, how it is retained and lost, how it is measured and overwritten, and whether these properties can be deduced from the structure of a model before its equations of motion are solved.

A material remembers when a write protocol alters its state, the writing perturbation is removed, and a later measurement still depends on that past intervention~\cite{HyperionCompanion}. This operational criterion covers a magnetic configuration kept behind an energy barrier, an internal difference of concentration relaxing across coupled compartments, and the shifted phase of a noisy limit-cycle oscillator. The criterion also shows why memory appears only when all parts of a physical description are specified together. The dynamical law decides whether a written displacement is amplified, transported or relaxed. The bath closure introduces thermal fluctuations and breaks time-reversal invariance. The driving protocol supplies the write perturbations and their temporal sequence. The measurement projection decides whether a stored pattern is resolved or averaged away into a vanishing global order parameter (Sec.~\ref{sec:construction}).

A history can record more than the identity of a single state. It can record the temporal order in which writes were applied, the compromise configuration that a write selects among several in a frustrated interaction network, or a distinction stored in degrees of freedom that are not observed. All of these records have the same form. Comparing two routes that should lead to the same result, two orders of operations or two paths around a loop, can expose a mismatch that neither route contains alone. Commutators, connection curvature and loop holonomy give the algebraic, the infinitesimal and the finite form of this comparison. Once it is clear which operators and which paths are compared, sensitivity to temporal sequence, geometric frustration and non-Markovian memory kernels fall into one description, and their physical actions remain distinct.

In time, write generators that do not commute displace the state along their commutator (the Lie bracket of the two vector fields). The order of operations is recorded whenever histories that share the same final write still settle into different states. On a spatial network, by contrast, local bond transformations may commute at a single site, but transport operations that shift states between sites need not commute. Their commutator defines the curvature of the bond connection, and composing them around a closed cycle gives the loop holonomy. When that loop holonomy admits no invariant state, no configuration can satisfy all bonds at once, and the cycle is frustrated. A third comparison arises under coarse-graining: the order of time evolution and projection onto the observed coordinates matters as well. Interchanging the two produces a Mori--Zwanzig memory kernel whenever the eliminated degrees of freedom are both excited by the observed channel and coupled back to it (Sec.~\ref{sec:record}). The stability of a record is separate from the way it was written: restoring forces, fluctuations and barriers decide how long it lasts and how it can be replaced (Secs.~\ref{sec:write}--\ref{sec:rewriting}).

Several properties of a memory can be read from the structure of a model before any trajectory is integrated: the bond transports decide frustration, the symmetries of the drift decide the kind of write, and the signs of the loops that close through the drive decide return-point memory. Continuation and reduction across published dynamical models and semiclassical hardware systems distinguish writes at the linear stage, writes at the nonlinear stage, shallow equilibrium exchange and multi-mode sweeps that have not settled (Secs.~\ref{sec:rules} and~\ref{sec:write-points}). The reflection transport structure of capillary colloidal rotors, transferred to arrays of in-plane point dipoles, predicts an orientational memory that the net magnetization does not show, and the calculation confirms it (Sec.~\ref{sec:carriers}). Collective Brownian dynamics show how a modulation of the coupling separates writing from retention (Sec.~\ref{sec:rewriting-carrier}). In a system of three coupled modes, two calculated fold instabilities define a coupling pulse that erases one stored pattern and preserves another (Sec.~\ref{sec:selective-erasure}). Full derivations and supplementary tests are given in the Supplementary Information (SI). Memory stored recursively in adapting material parameters is developed in a companion study~\cite{SelfImprovementCompanion}.
\section{Memory as a property of a realization}
\label{sec:construction}

\subsection{Components of a realization and their roles}

Whether a material remembers depends on its state space, its bath and the quantity that is measured, as well as on its equation of motion. A complete physical model therefore names all of these parts and is written as the realization~\cite{HyperionCompanion}
\begin{equation}
I_{\mathrm{real}} = \bigl((\Omega, \Xi); C, R, P; A\bigr).
\label{eq:identity}
\end{equation}
Here the dynamical operation $\Omega$ acts on the state carrier $\Xi$, the manifold or configuration space of admissible states. The closure $C$ specifies the boundary conditions, the conservation laws and the eliminated bath degrees of freedom, which supply thermal noise and dissipation. The observable $R$ defines the measured projection of the state. The protocol $P$ specifies how the system is prepared and driven in time. The assignment $A$ names the physical substrate and its material parameters. Together these components decide which distinction is written, which motions can erase it and which measurement can recover it.

The pitchfork normal form of Sec.~\ref{sec:write} is a single differential equation. As one control parameter and the noise intensity are varied, the same equation relaxes a written sign rapidly, slows its decay critically without retaining it, or preserves it across many relaxation times. The parameter regime and the bath closure decide which of these regimes occurs; the bare differential form does not. Rate-independent hysteresis models, such as the Preisach and play--stop operators~\cite{Preisach1935,Mayergoyz1991}, appear to place memory directly in the operator, because by definition their output depends on past turning points of the input. This placement is a feature of the coarse-grained description. The Preisach operator is physically equivalent to a superposition of elementary bistable relays~\cite{Mayergoyz1991}, so its memory is that of metastable elements retained behind energy barriers large compared with thermal noise.

\begin{figure}[!tbp]
\centering
\includegraphics[width=\linewidth]{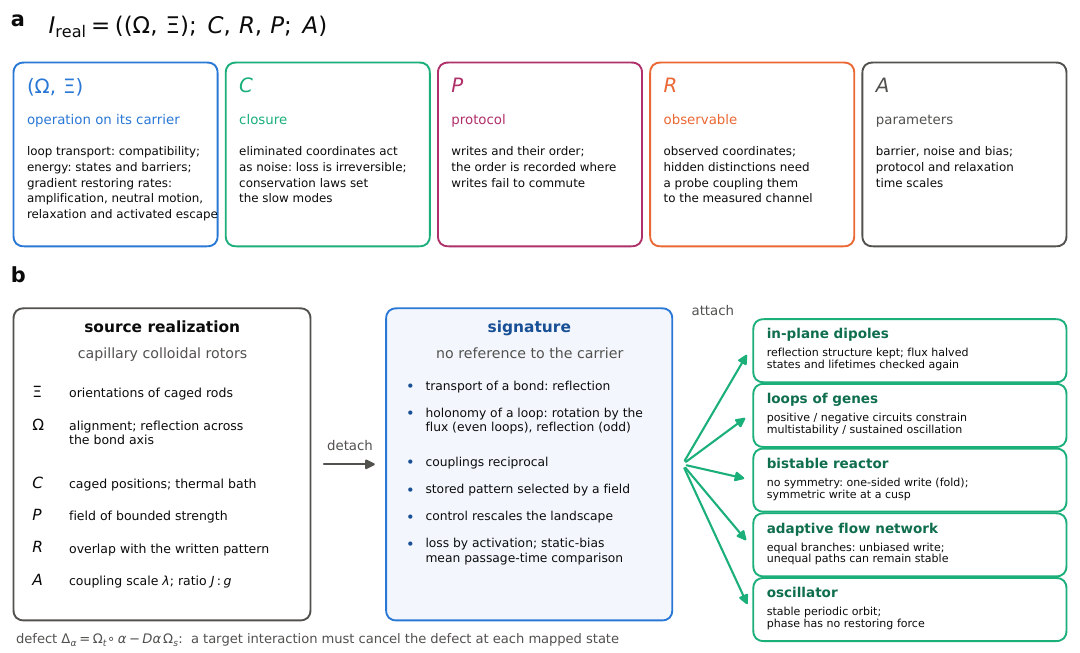}
\caption{\textbf{Memory as a property of a realization.}
\textbf{a}, The components of a realization, Eq.~\eqref{eq:identity}, and the aspect of memory that each component fixes.
\textbf{b}, Construction on another carrier. The structural signature of the colloidal memory consists of the invariants of the realization that do not depend on the substrate. It is detached from the network of capillary rotors and transferred to other physical carriers. On each target carrier the construction preserves part of the signature and modifies the rest. The compatibility defect of Eq.~\eqref{eq:defect} gives the physical terms that the target substrate must supply.}
\label{fig:construction}
\end{figure}

At every state on $\Xi$, the operation $\Omega$ sets the local restoring rates against displacements along each coordinate (Fig.~\ref{fig:construction}a). For an overdamped gradient flow with unit mobility, these rates are the eigenvalues $\kappa$ of the Hessian of the potential $V$. Negative curvature amplifies an initial bias away from an instability ridge. Positive curvature restores the coordinate toward a minimum. A vanishing quadratic curvature gives a flat direction when the higher-order restoring terms also vanish along it (Sec.~\ref{sec:signature}). Non-gradient flows need not admit a scalar potential. Their restoring rates are the eigenvalues of the Jacobian or, around a limit cycle, the Floquet multipliers.

On an interaction network, the coupling terms in $\Omega$ also define how a state is transported across each bond. Composing these bond transports around closed cycles decides whether the interaction preferences are mutually compatible or geometrically frustrated. Counting the resulting metastable states requires the full interaction energy (Sec.~\ref{sec:rules}).

The closure $C$ is the physical origin of forgetting. It eliminates the microscopic bath coordinates in favour of thermal noise and dissipation, and so breaks the microscopic reversibility under which fine-grained information would be conserved indefinitely. The protocol $P$ supplies the external drives and their order in time. The observable $R$ determines which coordinates are projected onto a measurement, whether by an external instrument or through the altered mechanical or chemical response of the material to a later drive. The material parameters in $A$ enter through dimensionless ratios that compare barrier heights, the effective write bias and the amplified signal with the thermal energy $k_BT$.

\subsection{Records of a history}
\label{sec:record}

An order record, a frustrated interaction network and a projected memory kernel store more than the identity of one selected state. All three arise where two pathways between descriptions fail to commute. For an order record, the two pathways apply the same write operations in opposite temporal sequences. For geometric frustration, they transport a state along alternative spatial paths that connect the same network elements. For a reduced description, they interchange time evolution with the elimination of unobserved coordinates. In each case the mismatch between the two pathways, the compatibility defect, carries a physical distinction that neither step contains on its own. The distinction becomes a physical record when it is written into the state, retained by the landscape and resolved by a later measurement.

Records of temporal order compare two pathways along a control protocol. A protocol history is an ordered sequence of write operations, free relaxations and measurements applied to a material. If the composite evolution did not depend on the order of its steps, the final state could reflect at most which operations had occurred, never their sequence. For continuously generated operations of durations $t_a$ and $t_b$, the commutator of their generators sets the leading order-dependent displacement,
\begin{equation}
e^{t_bL_b}e^{t_aL_a} - e^{t_aL_a}e^{t_bL_b} = t_at_b\,[L_b, L_a] + \dots
\label{eq:order-commutator}
\end{equation}
For deterministic flows, $L_a$ and $L_b$ act on observables as first-order differential operators. Their commutator $[L_b, L_a]$ is again a first-order differential operator, the Lie bracket of the corresponding vector fields~\cite{Arnold1989}, and it gives the leading order-dependent displacement of the state. When the operations are invertible, applying $L_a$, then $L_b$, then the inverse of $L_a$ and then the inverse of $L_b$ closes a loop in control space. The loop displaces the state along $[L_b, L_a]$, with a sign fixed by the orientation of the loop, and this displacement is the holonomy of the control cycle. Once the landscape stabilizes the displacement and the observable $R$ resolves it, the displacement records the temporal order of the writes.

Transport in space makes the same comparison. On a discrete network or lattice, the covariant shift $\mathcal S_k f(x)=g_k(x)f(x+\hat k)$ combines a shift between sites with the bond transformation $g_k(x)$. Its commutator compares the two routes across a plaquette, and it differs from the commutator $[g_k(x),g_l(x)]$ taken at a single site. For example, three antiferromagnetic sign inversions around a triangle commute pairwise, since each is a scalar multiplication by $-1$, yet their loop product is $(-1)^3=-1$. SI Sec.~\ref{si-sec:si-transport} gives the discrete calculation, including the orientation of transport.

In the continuum limit, spatial transport is governed by a smooth connection $\mathcal A_k$~\cite{KobayashiNomizu1996}, and the covariant derivatives $\nabla_k = \partial_k + \mathcal A_k$ combine a spatial shift with an internal transformation. They satisfy
\begin{equation}
[\nabla_k,\nabla_l]=F_{kl},\qquad
F_{kl}=\partial_k \mathcal A_l-\partial_l \mathcal A_k+[\mathcal A_k,\mathcal A_l].
\label{eq:connection-curvature}
\end{equation}
The curvature $F_{kl}$ measures the local mismatch of transport. Even for an Abelian connection, bonds that vary in space can give $\partial_k \mathcal A_l\ne\partial_l \mathcal A_k$ and a nonzero curvature. With the forward parallel-transport convention $\mathrm d\psi/\mathrm dt=-\mathcal A_k\,\dot x^k\psi$, the finite holonomy around a closed path $\gamma$ is
\begin{equation}
H_\gamma=\mathcal P\exp\!\left[-\oint_\gamma \mathcal A_k\,\mathrm dx^k\right],
\label{eq:finite-holonomy}
\end{equation}
where $\mathcal P$ orders the transformations along the path. For an infinitesimal positively oriented plaquette this gives $H_{\partial S}=\mathbb I-F_{kl}\delta S+o(\delta S)$. Reversing the orientation of the plaquette reverses the sign of the leading term. Finite loops require the full ordered transport. Noncontractible cycles can retain a global holonomy even where the local curvature vanishes.

This transport curvature compares spatial pathways. It is distinct from the Hessian curvature $\kappa$ of an energy landscape, which measures local restoring forces. When a continuous symmetry moves an equilibrium along a smooth orbit of distinct equilibria, the tangent to that orbit feels no deterministic restoring force ($\kappa = 0$). Thermal noise projected along the orbit then makes a written value diffuse. An exact conservation law acts differently. If both the deterministic drift and the noise preserve a global total, that total never diffuses. Spatial redistributions at fixed total relax through slow diffusive modes whose rates vanish with the wavenumber (Sec.~\ref{sec:time}).

Spatial holonomy, temporal order commutators and projected memory kernels share the comparison of pathways, but their physical consequences depend on the spaces and states on which they act~\cite{HyperionCompanion}. On an interaction network, the loop holonomy measures the mismatch of transport around a cycle. Geometric frustration occurs when no state is left unchanged by the loop holonomy, that is, when the holonomy has no fixed point. For $\mathbb{Z}_2$ spin inversions acting on nonzero spins, or for $SO(2)$ rotations acting on phases, only the identity map leaves a state unchanged, so any loop holonomy other than the identity frustrates the cycle.

An $O(2)$ reflection behaves differently. It leaves every orientation along its mirror axis unchanged, so a cycle with a nontrivial reflection holonomy can remain completely unfrustrated (Sec.~\ref{sec:rules}). On a network with several cycles, compatibility requires that all loop holonomies, parallel-transported to a common base node, share a common fixed state. Separate fixed states on the individual cycles are not enough. For interactions whose bond minimum defines an invertible transport, failure of this fixed-point condition means that the bond minima cannot all be reached at once.

In the capillary rotor networks of Sec.~\ref{sec:rules}, an even number of bond reflections composes to a rotation by an angle $\Phi_{\gamma}$ set by the bond geometry. This angle is analogous to the magnetic flux threading a ring of Josephson junctions~\cite{TeitelJayaprakash1983}, and the resulting compromise states depend on both this mismatch and the coupling strengths.

In directed regulatory circuits, the signs of the bonds constrain the feedback loops instead. Thomas's rules on the signs of feedback cycles then constrain the attractors (Sec.~\ref{sec:loops}).

Spatial frustration, temporal non-commutativity and the projection of coordinates act on different components of $I_{\mathrm{real}}$, so each can operate without the others. An unfrustrated ring of bistable elements records the temporal order of three writes (Sec.~\ref{sec:reading}). Two coupled relaxing coordinates generate a memory kernel without spatial cycles or multiple writes.

Unobserved coordinates can be eliminated by a projection operator $\Pi$ onto the retained subspace, with complementary projection $\Pi^{\perp} = \mathbb{I} - \Pi$. The reduced evolution then fails to compose as a Markovian semigroup, because evolving over $t+s$ differs from evolving over $t$, projecting, and evolving over $s$. For a time-independent generator $\mathcal{L}$, this mismatch of pathways produces the Mori--Zwanzig convolution kernel~\cite{Mori1965,Zwanzig1961,Zwanzig2001}
\begin{equation}
K(t) = -\Pi\,[\mathcal{L}, \Pi]\; e^{\Pi^{\perp}\mathcal{L}\Pi^{\perp}t}\; [\mathcal{L}, \Pi]\,\Pi
\label{eq:kernel-projection}
\end{equation}
(SI Sec.~\ref{si-sec:si-kernels}), which is exact for linear state dynamics or for linear projections of operators. Read from right to left, Eq.~\eqref{eq:kernel-projection} requires three steps. The retained coordinates must excite the orthogonal sector, $\Pi^{\perp}\mathcal{L}\Pi = [\mathcal{L}, \Pi]\Pi$. The excitation must propagate within the orthogonal sector under $e^{\Pi^{\perp}\mathcal{L}\Pi^{\perp}t}$. It must then feed back to the retained sector, $\Pi\mathcal{L}\Pi^{\perp} = -\Pi[\mathcal{L}, \Pi]$. When $[\mathcal{L}, \Pi] = 0$, the retained subspace decouples and $K(t)$ vanishes identically. If the coupling acts in one direction only, $[\mathcal{L}, \Pi] \neq 0$ can transmit an initial preparation from the hidden sector without generating a feedback kernel.

In linear dynamics, a single relaxing hidden mode gives an exponential kernel. A diagonalizable finite set of modes gives a sum of exponentials, with polynomial prefactors for defective Jordan blocks. A hierarchical spectrum of relaxations produces the stretched exponential characteristic of physical gels (SI Sec.~\ref{si-sec:si-sources}). Nonlinear projection formalisms and two-time kernels $K(t, t')$ extend this structure to thermal baths and to non-stationary aging media such as structural and spin glasses~\cite{Zwanzig1973,Meyer2017NonStationary,Bouchaud1992Aging,Cugliandolo1993Aging}.

Conversely, any strictly proper rational linear response can be generated by a minimal Markovian linear system whose dimension equals its Hankel rank~\cite{HoKalman1966}. When all poles of the response have negative real parts, this realization consists of relaxing modes; rationality alone does not make them stable. Any causal time-invariant nonlinear operator with fading memory on bounded inputs can be approximated by a finite Volterra series, or by a bank of linear modes followed by a static nonlinear output function~\cite{BoydChua1985}. Memory need not take the form of a kernel, however. A distinction carried directly by an observed bistable or neutral coordinate obeys a Markovian equation for that coordinate alone.

Suppose two preparations differ by $\delta q$ entirely within the eliminated subspace $\Pi^{\perp}\Xi$. A linear probe evolution $U_{\mathrm{probe}}(t)$ distinguishes them in the measured channel if and only if $\Pi U_{\mathrm{probe}}(t) \Pi^{\perp} \delta q \neq 0$ (Sec.~\ref{sec:reading}). For a time-independent probe generator, a nonzero $\Pi L_{\mathrm{probe}} \Pi^{\perp} \delta q$ gives a signal linear in time. If this term vanishes, the distinction can still reach the observable through intermediate coordinates, and the first nonzero $\Pi L_{\mathrm{probe}}^n \Pi^{\perp}\delta q$ gives a signal of order $t^n$.

A mismatch of pathways is required to generate an order record, a frustrated compromise or a memory kernel, but it does not by itself preserve the distinction. A register whose assignments overwrite the state is non-commutative, yet it retains only the final input. A frustrated cycle need not have several isolated minima. A path-dependent distinction becomes a retained record only if the landscape along the coordinate that carries it preserves it, through a neutral symmetry direction, a slow relaxing mode or an energy barrier between metastable basins. Pairing the origin of the stored distinction with the dynamical law that erases it classifies the mechanisms of material memory (Table~\ref{tab:classes}).

\subsection{Classes of memory mechanisms}
\label{sec:classes}

Table~\ref{tab:classes} classifies physical memory mechanisms by three properties. The first is the distinction that the state records. It is generated either by one of the mismatches of pathways of Sec.~\ref{sec:record} or by direct selection among multistable wells. The second is the dynamical law of loss that governs the coordinate carrying the record: exponential relaxation, algebraic diffusion or Kramers activation (Sec.~\ref{sec:loss-laws}). The third, the bifurcation normal form at the instability where the state is selected, specifies how the record is written. It is written symmetrically across a pitchfork, in one direction across a fold threshold, or into an oscillation across a Hopf bifurcation, whose phase can later store a shift (Sec.~\ref{sec:rules}). A single material can carry distinct records in different coordinates, so the seven rows of Table~\ref{tab:classes} are overlapping classes of mechanism and not mutually exclusive compartments.

\begin{table}[!tbp]
\centering
\footnotesize
\setlength{\tabcolsep}{3pt}
\renewcommand{\arraystretch}{1.15}
\begin{tabular}{>{\raggedright\arraybackslash}p{0.14\linewidth}>{\raggedright\arraybackslash}p{0.25\linewidth}>{\raggedright\arraybackslash}p{0.15\linewidth}>{\raggedright\arraybackslash}p{0.15\linewidth}>{\raggedright\arraybackslash}p{0.23\linewidth}}
\toprule
Record & Structural origin & Write & Law of loss & Examples \\
\midrule
One of a few stable states & A non-convex energy of one element, or a positive feedback & Pitchfork (symmetric) or fold (one-sided) & Activation & Bistable beams and folds~\cite{Silverberg2014Origami,Florijn2014Metamaterials,Chen2021Reprogrammable}; genetic toggle~\cite{Gardner2000Toggle}; magnetic grains~\cite{Kryder2008HAMR}; filamentary memristors~\cite{Yang2013Memristor} \\
One of many compromise states & Frustration: loops whose holonomy leaves no state unchanged, together with quenched disorder & A field that removes the stored state; a coupling raised during the write & Activation & Frustrated magnets~\cite{Moessner2006} and artificial spin ice~\cite{Wang2006SpinIce}; capillary colloids~\cite{BaulinColloidMemory}; dipole arrays (Sec.~\ref{sec:carriers}) \\
Extreme values and turning points of a scalar drive & Elements that switch at thresholds of the drive; exact return to a turning point when no loop through the drive is frustrated (rule R8, Sec.~\ref{sec:return-point}) & Thresholds of the drive & Activation, with barriers large compared with the noise & Ferromagnets~\cite{Barkhausen1919,Preisach1935,Sethna1993Hysteresis}; sheared suspensions~\cite{Corte2008Reversible,Keim2011Memory,Paulsen2014Memory} \\
Order of operations & Writes that fail to commute on the coordinates that retain the state, so that the final state depends on their order; a holonomy when the writes can be reversed & A sequence whose order matters; a common final input tests more than the last write & The law of the retaining coordinates & Ring of bistable coordinates (Sec.~\ref{sec:reading})~\cite{SequenceCompanion}; zero-temperature Ising model~\cite{Croce2026}; mechanical metamaterials~\cite{Kwakernaak2023} \\
Recent history, fading with time & Slow coordinates coupled to the observed ones, which give a memory kernel when eliminated & Any excitation of the slow coordinates & Relaxation at the rates of the slow coordinates, slower with age in glasses & Hydrogels~\cite{Yu2020HydrogelMemory,Zhao2025GelHierarchy}; aging and the Kovacs effect~\cite{Kovacs1979,Bouchaud1992Aging,Cugliandolo1993Aging}; exchange between compartments (Sec.~\ref{sec:reading})~\cite{HyperionCompanion} \\
An analogue value or a phase & A neutral direction with noise acting along it & A force along the flat direction; a change of frequency & Diffusion; on compact carriers the value eventually mixes exponentially & Phase of a genetic oscillator~\cite{Elowitz2000Repressilator}; charge in linear-drift memristor models~\cite{Chua1971Memristor,Strukov2008Memristor} \\
A plastic deformation & Translational holonomy of circuits in the lattice: the Burgers vector of dislocations & Plastic flow & Activation of the motion and annihilation of defects & Dislocations in crystals~\cite{Nye1953,Ashby1970} \\
\bottomrule
\end{tabular}
\caption{\textbf{Classes of memory mechanisms.} Each class is defined by three properties. The first is what it records, which follows from a realization of the principle of Sec.~\ref{sec:record} or, in the first row, from multistability alone. The second is the law of loss of the coordinate that carries the record (Sec.~\ref{sec:loss-laws}). The third is the kind of write (Sec.~\ref{sec:rules}). The rows are not exclusive, since one material can carry several records in different coordinates. SI Table~\ref{si-tab:sources} lists the physical sources within each law and the controls that remove them.}
\label{tab:classes}
\end{table}

Materials of entirely different chemical composition share a memory class when their interaction structure, state space and dynamics of loss coincide. System size, spatial dimension and noise intensity enter through dimensionless ratios: the write number of Eq.~\eqref{eq:write-law}, the ratio of barrier to noise, and the effective energy bias in units of $k_BT$. One material can host several classes at once in different coordinates. A ferromagnet records the turning points of an applied field through its pinned domain walls, and it chooses the overall sign of its magnetization at a pitchfork when it is cooled through its Curie point in a weak field. Two families of memory that are often treated separately also belong to Table~\ref{tab:classes}. The memories of athermal hysteretic elements, sheared suspensions and mechanical metamaterials are activated retention in the limit where the energy barriers far exceed the thermal fluctuations, and they disappear once the noise becomes comparable to the barriers. Topological memories combine two rows of the table. The holonomy of a circuit around a defect determines what is recorded, and the barriers against motion and annihilation of the defect determine how long the record survives. Every class admits an experimental control that removes the memory and leaves the material intact, and this control identifies the mechanism in practice (SI Table~\ref{si-tab:sources}).

The fifth class, fading memory, forgets the distant past asymptotically. For causal, time-invariant linear responses, fading memory is represented by a convolution kernel. On the bounded input spaces of Boyd and Chua, any continuous time-invariant nonlinear operator with fading memory can be approximated by a finite Volterra series~\cite{BoydChua1985}. An operator that records the running maximum of its input retains arbitrarily old extrema and so lies outside fading memory. So do noiseless neutral directions and basins separated by impenetrable barriers. At finite noise these records also decay eventually, through diffusion or Kramers activation, on timescales fixed by the retaining coordinates.

Records can be nested in time, in scale and in level. The three kinds of nesting correspond to three components of a realization, Eq.~\eqref{eq:identity}. The protocol carries nesting in time. The observable and the coordinates that the closure eliminates carry nesting in scale. The parameters carry nesting in level, since they become the carrier of the level above. The turning points of a drive form a nested record. An excursion that returns erases the turning points inside it and keeps those that enclose it (Sec.~\ref{sec:return-point})~\cite{Sethna1993Hysteresis,Mayergoyz1991}. In the language of computation the record is a stack: a turning point is added when the drive reverses and removed when the drive passes it again.

A record can also lie beneath a global order parameter that averages it away. Alignment alone gives the colloidal layer (anisotropic colloids at a fluid interface whose orientations interact through nematic alignment and a bond-directional capillary coupling; Sec.~\ref{sec:construction-step}) a common director. The director is a flat direction that stores one angle and loses it as the director wanders (Sec.~\ref{sec:rules}). On a regular lattice the capillary preferences of each particle cancel for any common orientation, and alignment still orders the layer. Positional disorder removes this cancellation particle by particle. Each particle acquires its own preferred axis, as in a random-anisotropy magnet~\cite{Garanin2022RandomAnisotropy}, the global order stays weak, and the written pattern resides in the orientations of the particles relative to their bonds. The two levels compete. Below a positional disorder of about a tenth of the particle spacing, part of the written state is carried by the director. Above it the written overlap decreases although the global order stays small~\cite{BaulinColloidMemory}. A probe of the global order alone then reports no memory, as a measurement of the net moment does for the dipole arrays of Sec.~\ref{sec:carriers}.

Nesting in scale can extend over many scales when the barrier against a rearrangement grows with its size. In spin glasses the barriers grow as a power of the size of the rearranged region~\cite{FisherHuse1988}. At a lower temperature the larger scales keep the configuration reached at a higher one while the smaller scales are rewritten. A spin glass cooled in steps therefore keeps a record of each step and recovers it on reheating~\cite{Jonason1998Memory,Bouchaud2001Microscope}. Relaxation rates distributed without a characteristic scale give logarithmic relaxation and aging~\cite{AmirOregImry2012}. At the critical disorder of the random-field Ising model, the avalanches by which the state changes have no characteristic size~\cite{SethnaDahmenMyers2001}. The energy of large rearrangements bounds the depth of nesting in scale. In two dimensions a network with random loop fluxes, the gauge glass, has no order at a finite temperature beyond a correlation length that diverges only as the temperature vanishes~\cite{KatzgraberYoung2002}. Positional disorder randomizes the loop fluxes of the capillary layer through Eq.~\eqref{eq:loop-flux} of Sec.~\ref{sec:rules}, and the depth of its nesting in scale is open to measurement. The levels need not be retained behind barriers. When a cascade of instabilities writes them one after another, each stage seeding the next and finer one, the write number of each stage sets the depth instead, and the cascade ends where noise or dissipation at the finest scale makes the selection random~\cite{StructuralComplexityCompanion}.

The parameters of a material can carry a memory one level above its states, and the parameters of the rule by which the material adjusts them can carry one a level higher still (Sec.~\ref{sec:discussion}). The same properties classify each level of such a hierarchy.

\subsection{Construction on another carrier}
\label{sec:construction-step}

A realization is mapped from a source carrier $\Xi_s$ to a target carrier $\Xi_t$ by a state-space map $\alpha: \Xi_s \to \Xi_t$ that sends source states to target states~\cite{HyperionCompanion} (Fig.~\ref{fig:construction}b). The target vector field $\Omega_t$ reproduces the source dynamics $\Omega_s$ if evolving on the source and then mapping to the target gives the same trajectory as mapping first and then evolving on the target. The difference between these two routes is the compatibility defect
\begin{equation}
\Delta_\alpha(x) = \Omega_t\bigl(\alpha(x)\bigr)-\partial\alpha(x)\,\Omega_s(x),
\label{eq:defect}
\end{equation}
where $\partial\alpha$ is the tangent map of $\alpha$. Where $\Delta_\alpha = 0$, the mapped source trajectories solve the target equation directly. Where $\Delta_\alpha \neq 0$, the target carrier must supply a compensating interaction or control force equal to $-\Delta_\alpha(x)$ at the target state $\alpha(x)$. Such a force exists on $\Xi_t$ when $\Delta_\alpha(x)$ takes the same value for all source states that map to the same target state and when admissible target interactions can produce it. Otherwise the target needs additional coordinates or a refined map. For stochastic dynamics, exact transfer requires the generators to intertwine on observables, so that both the transformed diffusion tensor and the It\^o drift correction match.

For linear maps, Eq.~\eqref{eq:defect} reduces to $\Delta = \Omega_t\alpha - \alpha\Omega_s$, and it contains all the comparisons of pathways of Sec.~\ref{sec:record} as special cases. Taking $\alpha = \Pi$ as the coarse-graining projection between $\Omega_s = \mathcal{L}$ and $\Omega_t = \Pi\mathcal{L}\Pi$ gives the return defect $\Delta = \Pi\mathcal{L}\Pi - \Pi\mathcal{L} = -\Pi\mathcal{L}\Pi^{\perp} = \Pi[\mathcal{L}, \Pi]$. Paired with the excitation commutator $\Pi^{\perp}\mathcal{L}\Pi$, it forms the Mori--Zwanzig kernel of Eq.~\eqref{eq:kernel-projection}. Taking $\Omega_s = \Omega_t = L_b$ and $\alpha = e^{t_a L_a}$ on a single carrier gives $\Delta = [L_b, e^{t_a L_a}]$, whose leading term $t_a[L_b, L_a]$ gives the order commutator of Eq.~\eqref{eq:order-commutator}. Comparing spatial transport pathways around a network cycle gives the loop holonomy. In every setting, a non-vanishing compatibility defect identifies what a reduced or transferred description omits and specifies the interaction or coordinate needed to complete it~\cite{HyperionCompanion}.

On a new carrier, the writing drive must couple to the coordinate intended to store the record. A force can displace a stable coordinate, a frequency pulse can shift the phase of an oscillator, and an instability can amplify a weak bias into a choice between basins. The restoring rates and the barriers then set retention. The compatibility defect $\Delta_\alpha$ identifies the interaction or additional coordinate needed to reproduce the specified source dynamics on the target.

A comparison across carriers tracks five structural properties of a realization: the bond-transport group and its loop holonomies, the reciprocity of the couplings, the equivariant symmetries of the drift, the structural role of the control parameter, and the law of loss. A given transfer may preserve part of this signature and modify the rest. If it preserves the bond-transport action and the reciprocity, the loop compatibility test keeps its algebraic form. The specific loop fluxes, the number of metastable states and the barrier heights depend on the full target energy and must be evaluated on the target carrier. Preserving a quantitative writing law further requires matching the reduced normal form, the projected bias and noise, and the separation of timescales from slower modes.

In a monolayer of capillary colloidal rotors~\cite{BaulinColloidMemory}, anisotropic particles are trapped at a fluid interface. Their centres remain caged by neighbours, while their in-plane orientations $\theta_i$ rotate on the $N$-torus modulo $\pi$. Freezing the particle centres over rotational timescales is part of the closure $C$. The interaction energy combines two terms. Nematic alignment transports an orientation unchanged across a bond. A quadrupolar capillary coupling (Eq.~\eqref{eq:compass-coupling}) reflects an orientation across the axis of the bond between the two particles. Composing these bond reflections around closed cycles gives the geometric loop compatibility conditions of Sec.~\ref{sec:rules}. The full interaction energy, scaled by the coupling strength $\lambda$, sets the metastable compromise basins and their Kramers barriers. A patterned external field writes a chosen angular configuration. The observable $R$ measures the connected overlap with that pattern, which remains hidden from the vanishing macroscopic nematic director. Transferring this memory to a new substrate requires comparing the bond-transport geometry and also the resulting energy landscape, the fluctuations and the observable (Sec.~\ref{sec:carriers}).
\section{Writing, retention, measurement and rewriting}
\label{sec:signature}

Restoring forces control local relaxation, while fluctuations spread the state along neutral coordinates or carry it across barriers. These motions separate the stability of a record from the way it was written. A barrier that protects a stored pattern also obstructs its replacement through the same transition path. A static bias can lower that obstacle. Alternatively, a material can be written while its barriers are low and stabilized afterwards. Comparing these operations requires the strength of the bias as well as the landscape and the region in which the state is stored. Section~\ref{sec:rewriting} develops this comparison for mean transition times.

\subsection{Writing at an instability}
\label{sec:write}

\begin{figure}[!tbp]
\centering
\includegraphics[width=\linewidth]{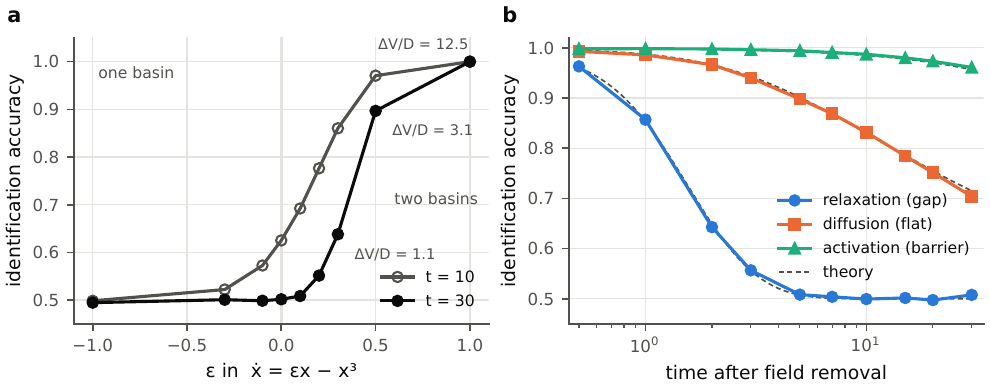}
\caption{\textbf{One operation on both sides of its write point, and the laws of loss.}
\textbf{a}, Accuracy with which a binary write is identified for the single operation $\dot{x} = \epsilon x - x^3$ at noise $D = 0.02$, measured at $t = 10$ and $t = 30$ after the field is removed. For $\epsilon < 0$ the write relaxes. At $\epsilon = 0$ critical slowing lengthens the transient but does not retain the write. For $\epsilon > 0$ the write is retained only as $\Delta V/D$ grows. Labels give $\Delta V/D$.
\textbf{b}, Accuracy after a binary write and removal of the field at noise $D = 0.05$, for relaxation ($\kappa = 1$), diffusion ($\kappa = 0$) and activation ($\Delta V = 0.25$). Dashed curves, predictions of Sec.~\ref{sec:loss-laws}. Each point is estimated from 4{,}000 independent Langevin realizations for each sign of the input.}
\label{fig:loss-laws}
\end{figure}

The simplest equation in which a state is both written and retained is the pitchfork normal form,
\begin{equation}
\dot x = \epsilon x - x^3 + \sqrt{2D}\,\xi(t),
\label{eq:pitchfork}
\end{equation}
with the Landau potential $V(x) = -\epsilon x^2/2 + x^4/4$. The operation is the same throughout this section; only the parameter $\epsilon$ and the noise intensity $D$ change. For $\epsilon < 0$ the potential has one minimum, and a small displacement relaxes deterministically at the rate $|\epsilon|$. At $\epsilon = 0$ the restoring force is $-x^3$, which is weak near the minimum. Without noise the relaxation is then algebraic, $x(t) = x_0/\sqrt{1 + 2x_0^2t}$. At finite $D$ the quartic potential confines a stationary distribution of width of order $D^{1/4}$, and the relaxation at long times is exponential, on a timescale of order $D^{-1/2}$. Critical slowing prolongs the transient, but in neither case is there a second stable state in which a written sign could be retained. For $\epsilon > 0$ two minima at $\pm\sqrt{\epsilon}$ are separated by a barrier $\Delta V = \epsilon^2/4$, and the state changes wells only by thermal activation. The two minima retain a written sign only when the barrier is large compared with the noise. At $\Delta V/D$ of order one the sign is lost within a few relaxation times (Fig.~\ref{fig:loss-laws}a; SI Sec.~\ref{si-sec:si-laws}). The one operation therefore produces relaxation, critical slowing and activated retention. It never produces a flat direction, which would require a continuous symmetry or a conserved quantity that this operation does not have.

For $\epsilon > 0$ the origin is a ridge, with curvature $V''(0) = -\epsilon$. A state released near the ridge at a small displacement $s$ is pushed away from it at the rate $\epsilon$, and its thermal spread is amplified at the same rate. In the linearized dynamics $x(t)\,e^{-\epsilon t}$ converges to $s + \eta$, with $\eta$ Gaussian of variance $D/\epsilon$. Only the noise that acts while the state is still near the ridge competes with the bias. Once the state has moved away, bias and noise are amplified together. In the linear approximation the information about the bias then approaches a plateau, before saturation and possible transitions between the wells set in. An unstable direction therefore writes a state; it does not add a further way of losing one (Sec.~\ref{sec:loss-laws}). The choice between the wells is made within about one amplification time, and the state ends in the well favoured by the bias with probability
\begin{equation}
p_{\mathrm{write}} = \Phi\bigl(\sqrt{W}\bigr), \qquad W = \frac{\epsilon s^2}{D},
\label{eq:write-law}
\end{equation}
where $\Phi$ is the standard normal distribution function. The write number $W$ compares the bias, amplified over one amplification time, with the noise accumulated in that time. When the state settles into a well, the amplified amplitude is reduced to its sign. The sign keeps a fraction of the information in the amplitude, from $2/\pi$ for a weak bias to all of it for a strong one ($\WriteKeptLow$ at $W = 1/16$ and $\WriteKeptHigh$ at $W = 4$). Simulations of Eq.~\eqref{eq:pitchfork} reproduce Eq.~\eqref{eq:write-law} to within $\WriteInstantMaxDev$ for $W$ between $1/16$ and 4 (Methods).

A bias can also act while the control parameter is swept through the bifurcation, $\epsilon = rt$, as when a material is cooled through a symmetry-breaking transition in a weak field $h$. Linearizing about the ridge and integrating from well before the bifurcation gives the probability of ending in the favoured well,
\begin{equation}
p_{\mathrm{sweep}} = \Phi\!\left(\frac{\pi^{1/4}\, h}{\sqrt{D}\, r^{1/4}}\right),
\label{eq:sweep-law}
\end{equation}
which is the sensitivity of a slowly swept bifurcation to a weak bias~\cite{KondepudiNelson1985,KondepudiMossMcClintock1986,KondepudiMossMcClintock1986Letter}. A sweep at rate $r$ keeps the state near the ridge for a time of order $r^{-1/2}$. The displacement produced by the bias grows in proportion to this time, and the thermal spread grows only as its square root, so slower sweeping improves the accuracy only as $r^{-1/4}$. Simulations at sweep rates from 0.02 to 2 give $\SweepSimRange$, against a predicted $\SweepTheory$ (Methods).

Equation~\eqref{eq:sweep-law} is the accuracy with which a force can be detected on a free coordinate. A coordinate without a restoring force that is observed for a time $T$ under a force $h$ is displaced by $hT$ and spread by $\sqrt{2DT}$, and no decision based on its trajectory identifies the sign of $h$ with a probability above $\Phi\bigl(h\sqrt{T/2D}\bigr)$. The linear stage of the sweep integrates the bias and the noise with the same Gaussian weight $w(t)=e^{-rt^2/2}$, and Eq.~\eqref{eq:sweep-law} is this limit for the window $T=(\int w\,\mathrm dt)^2/\int w^2\,\mathrm dt = 2\sqrt{\pi}\,r^{-1/2}$. A bias kept on at a fixed ridge has the weight $e^{-\epsilon t}$, the window $2/\epsilon$ and the accuracy $\Phi\bigl(h/\sqrt{\epsilon D}\bigr)$. An instability therefore writes as an ideal detector of the bias would after observing the free coordinate for the time during which the restoring force is absent. In that time the bias does the work $h^2T = 2z^2D$ on the free coordinate, where $z$ is the argument of $\Phi$: about $3.3\,k_BT$ for an accuracy of 0.9 and $10.8\,k_BT$ for 0.99. This is the cost of a write at an instability, and it contains no property of the wells that form afterwards.

The accuracy does not depend on how deep the wells become afterwards, provided the noise is weak enough that the state does not cross back over the ridge while the barrier grows. Writing and retention therefore take place in different parts of the same landscape. The ridge decides the state, and the wells retain it. This separation allows a weak bias to rewrite a state that is retained behind a high barrier (Sec.~\ref{sec:rewriting}).

The ridge decides the state only if the choice is complete while the deviation still grows linearly. This condition can fail in two ways. When a term of higher order, or noise that is still comparable to the deviation, acts after the linear stage, the nonlinear dynamics completes the choice, and the probability of the favoured well departs from Eq.~\eqref{eq:sweep-law}. When the wells remain shallow after the sweep, the noise keeps exchanging the state between them. Their occupation is then set by the balance of the wells and not by the bias at the ridge; this is an equilibrium selection and not a write. The three cases are distinguished by continuing the sweep beyond the linear stage and then keeping the control at its final value for as long as the sweep lasted. A write made in the linear stage follows Eq.~\eqref{eq:sweep-law}. When the law fails, this interval distinguishes the other two cases. A write completed in the nonlinear stage keeps its state, whereas in an equilibrium selection more than a tenth of the trajectories change state during the interval (Sec.~\ref{sec:write-points}; Methods).

\subsection{Retention and loss: one expression and three regimes}
\label{sec:loss-laws}

After the write is removed, the coordinate $q \in \Xi$ that carries it evolves in contact with a thermal bath,
\begin{equation}
\mathrm{d}q = -V'(q)\,\mathrm{d}t + \sqrt{2D}\,\mathrm{d}B_t,
\label{eq:langevin}
\end{equation}
where time is measured with the mobility set to one, so that $V$ is the potential energy and $D=k_BT$ in these units. In physical time the drift is $-\mu V'$ and the diffusion coefficient is $\mu k_BT$, with mobility $\mu=1/\zeta$ and friction coefficient $\zeta$. Retention is measured by the mutual information $I(Y;q(t))$ between the write and the later state or, for a binary write, by the probability that the sign of $q(t)$ identifies the written sign. Within a region the local potential sets the relaxation. Between regions the barriers and the noise set the transitions.

With $V''(q) = \kappa$ of either sign, a written displacement drawn from a Gaussian of variance $s^2$ is multiplied by $e^{-\kappa t}$ at time $t$, and it is accompanied by independent Gaussian noise of variance $D(1 - e^{-2\kappa t})/\kappa$, so that
\begin{equation}
I(t) = \frac12\ln\left[1 + \frac{\kappa s^2/D}{e^{2\kappa t} - 1}\right].
\label{eq:information-any-kappa}
\end{equation}
For $\kappa>0$ the coordinate relaxes towards a minimum with stationary variance $\sigma^2=D/\kappa$, and $I(t)\simeq(s^2/2\sigma^2)e^{-2\kappa t}$. The information decays at twice the mechanical relaxation rate because it is quadratic in the surviving displacement.

On a flat, non-compact direction the limit $\kappa\to0$ gives variance $2Dt$ and $I(t)=\tfrac12\ln(1+s^2/2Dt)\simeq s^2/(4Dt)$. A finite number of independent flat coordinates with finite write covariance gives the same leading exponent. On a compact carrier this law describes the early spreading, before mixing becomes exponential (Sec.~\ref{sec:carriers}). An exactly conserved total is not described by this law, because noise that preserves the total cannot spread it.

For $\kappa<0$ the linear expression tends to $\tfrac12\ln(1+W)$, with $W=|\kappa|s^2/D$. The amplification and the accumulated noise then grow at the same rate, and their ratio, the signal-to-noise ratio, stays finite. In the full nonlinear write this stage ends at saturation, and the wells and their barriers decide retention afterwards.

When the written states are wells separated by a barrier $\Delta V$, the state first relaxes within its well and then crosses the barrier only rarely, at the Kramers rate~\cite{Kramers1940,Hanggi1990}
\begin{equation}
r_K = \frac{\sqrt{V''(q^*)\,|V''(q_b)|}}{2\pi}\,e^{-\Delta V/D},
\label{eq:kramers}
\end{equation}
where $q^*$ is the minimum and $q_b$ the saddle. Between the relaxation within the well and the first crossings, $I(t)$ stays on a plateau. The plateau equals the entropy of the write when the write places the state in the intended well without error. For two symmetric wells written with equal probability, the populations relax as $p(t) = \frac12(1 - e^{-2r_Kt})$, and
\begin{equation}
I(t) = \ln 2 - H_{\mathrm b}\bigl(p(t)\bigr) \;\simeq\; \frac12\,e^{-4r_Kt},
\label{eq:activation-law}
\end{equation}
where $H_{\mathrm b}$ is the binary entropy function. The three laws of loss are relaxation, diffusion and activation (Laws~1--3). The remaining sign of the curvature, $\kappa < 0$, writes a state instead of losing one. Each law sets the time over which a written distinction remains measurable: $1/\kappa$ for relaxation, $s^2/D$ for diffusion and $1/r_K$ for activation. Only the last grows exponentially with a property of the landscape, the ratio of the barrier to the noise, and this is why durable memories are retained behind barriers. Simulations of binary writes follow the three laws (Fig.~\ref{fig:loss-laws}b; SI Sec.~\ref{si-sec:si-laws}).

The three laws combine into a lifetime that can be read from the equations of a model with many coordinates and no potential, without integrating trajectories. At a stable state $q$ with Jacobian $\mathcal J$ and noise matrix $D$, the spread of the linearized motion is the covariance $\Sigma$ that solves $\mathcal J\Sigma+\Sigma\mathcal J^{\mathsf T}+2D=0$ (Law~1). The distance to the nearest unstable fixed point, or half the distance to another stable state, measured in this spread gives the exponent $S=\tfrac14\min_e\tfrac12 d^{\mathsf T}\Sigma^{-1}d$, where the minimum runs over the neighbouring fixed points $e$ and $d$ is the displacement to them. The exponent equals $\Delta V/D$ for the pitchfork normal form, Eq.~\eqref{eq:pitchfork}, and estimates it for other barrier shapes, and the rate $|\mathrm{Re}\,\lambda|\,e^{-S}/2\pi$, with $\lambda$ the slowest eigenvalue, takes the place of the Kramers rate (Law~3). A state with a zero eigenvalue is treated as lying on a line of states, without a barrier; noise spreads it along the line (Law~2). The rule was set on nine pitchfork models of the 134-model write-point survey (Sec.~\ref{sec:write-points}) and tested on the 31 others with two or more stable states, from biology, epidemiology, ecology, quantum hardware, fluid dynamics and cosmology (Methods). On these 31 models this lifetime classified all 27 states that have a barrier correctly as kept or lost, whereas the model records, which keep a stable state unless their own hold check shows a loss, gave 19; over all states, leaving out those from which most trajectories diverge, it was right for 65 of 75, against 23 (SI Sec.~\ref{si-sec:si-lifetime}). Its ten misclassified states belong to three epidemic models whose states lie on a line of disease-free states: noise moves the state along the line, but the measured quantity, whether the disease is present, does not change, and the record is kept. Drift along a flat direction erases a memory only when the observable varies along it, as it does under the common rotation of the rotors in Sec.~\ref{sec:flat}.

Law~1 has a general counterpart for any landscape with a stationary distribution. For overdamped dynamics with detailed balance, a normalizable stationary distribution $\pi \propto e^{-V/D}$ and a logarithmic Sobolev inequality, the information about a write cannot exceed the mean relative entropy $\mathcal{D}$ of the conditional states with respect to $\pi$, and this relative entropy decays at least as fast as $e^{-2\rho t}$:
\begin{equation}
I\bigl(Y; q(t)\bigr) \le \mathbb{E}_Y\, \mathcal{D}\bigl(p_t(\cdot\,|\,Y)\,\|\,\pi\bigr) \le e^{-2\rho t}\; \mathbb{E}_Y\, \mathcal{D}\bigl(p_0(\cdot\,|\,Y)\,\|\,\pi\bigr),
\label{eq:lsi-bound}
\end{equation}
where $\rho$ is the logarithmic Sobolev constant of the dynamics~\cite{Gross1975,BakryEmery1985}. For a smooth complete carrier without boundary, a uniform positive lower bound on $D\,\mathrm{Ric}+\mathrm{Hess}\,V$ gives a lower bound on $\rho$~\cite{BakryEmery1985,OttoVillani2000}; a carrier with boundaries requires compatible conditions at the boundary. A confining quadratic potential on a flat carrier attains $\rho=\kappa$. Unconfined Brownian motion has no normalizable stationary distribution, so its algebraic loss lies outside this equilibrium inequality. Diffusion on a compact carrier, by contrast, approaches its stationary measure exponentially. In a metastable landscape $\rho$ can be exponentially small~\cite{MenzSchlichting2014}, because the exchange between wells is slow. Related entropy inequalities apply to suitable Markov chains~\cite{DiaconisSaloffCoste1996,ErbarMaas2012}, to free-energy gradient flows~\cite{OttoVillani2000} and to quantum Markov semigroups~\cite{KastoryanoTemme2013,CarlenMaas2017} (SI Sec.~\ref{si-sec:si-bound}).

\subsection{Time and space in a realization}
\label{sec:time}

A closed Hamiltonian system conserves information, even when a magnetic field breaks its time-reversal symmetry. Information decreases because the closure discards a continuum of degrees of freedom, and their influence on the retained coordinates enters as noise uncorrelated with the write. When the closure is Markovian, the mutual information between the write and the state cannot increase~\cite{CoverThomas2006}, and Eq.~\eqref{eq:lsi-bound} bounds the rate at which it decreases. A bath with a memory of its own can return information to the retained coordinates for as long as its memory kernel lasts. Time also enters Eq.~\eqref{eq:identity} through the order in which a protocol applies its operations (Sec.~\ref{sec:reading}). It can also be a coordinate of the carrier: an operation with a stable limit cycle retains the phase along the cycle as a flat direction (Sec.~\ref{sec:carriers}).

For a conserved diffusive density, a nonzero Fourier mode relaxes at the rate $\kappa(k)=Mk^2$~\cite{HohenbergHalperin1977}. For the class of write spectra considered in SI Sec.~\ref{si-sec:si-fields}, a write whose spectral weight is proportional to $k^{2n}$ near zero gives a decaying contribution proportional to $t^{-(d+2n)/2}$ in $d$ dimensions. Adding material locally also changes the conserved zero mode, which retains that total indefinitely. Moving material between neighbouring sites contributes nothing to the zero mode and gives $n=1$. These are scalings of the continuum at long times, before a finite system resolves its smallest nonzero wavenumber. A positive spectral gap instead produces exponential loss. In this example conservation enforces gapless diffusive modes, but conservation is not the only source of slow modes: a non-conserved field at a critical point can also be gapless.

\subsection{Measurement and the order of writes}
\label{sec:reading}

A stored distinction can reside entirely in coordinates that the usual observable does not contain, and it can then be measured only with a suitable probe. Three compartments of equal volume that exchange a conserved species reciprocally give an exact example~\cite{HyperionCompanion}. Material is placed in one compartment, and two exchange operations are applied in either order. Both orders return the same population to that compartment, and relaxation under the symmetric exchange never reveals which order was applied. The reason is a symmetry. For reciprocal exchange the difference between the two orders is an antisymmetric matrix, and its diagonal, which gives the change of the returned population, vanishes. The order is still stored, in the imbalance between the other two compartments. A later exchange that favours one of them converts this imbalance into a difference of the observed population, and this difference passes through a maximum at a time set by the exchange rates (SI Sec.~\ref{si-sec:si-compartments}). Whether the order can be measured therefore depends on the probe as well as on the delay.

The laws of Sec.~\ref{sec:loss-laws} describe how long a distinction survives in the coordinates that carry it. A measurement requires in addition a probe whose generator couples those coordinates to the observable. The returned populations are equal because the exchange is reciprocal between compartments of equal volume and the measured quantity is a return probability; writes that are not reciprocal, such as writes by applied fields, generally change the returned population as well.

A device that overwrites its state, $W_A(x) = A$ and $W_B(x) = B$, has $W_BW_A \ne W_AW_B$ and still retains only its last input, so writes that fail to commute are not sufficient for a record of their order. With two writes this cannot be detected, because the order and the identity of the last write are then the same statement. A record of the order requires that orderings ending in the same write produce distinct measurements,
\begin{equation}
R\bigl[(W_A \circ W_C \circ W_B)(q_0)\bigr] \ne R\bigl[(W_A \circ W_B \circ W_C)(q_0)\bigr],
\label{eq:sequence-criterion}
\end{equation}
evaluated after relaxation without a field. If the final state were determined only by the identity of the final write, the six orderings of three writes would end in at most three states. In a ring of eight bistable coordinates coupled by a discrete Laplacian, a system without frustration, three writes that address the same coordinates end in \OrderStates\ states. For each of the \OrderClassesSplit\ final writes, the two orderings that share it end in different states and respond differently to one weak probe~\cite{SequenceCompanion} (SI Sec.~\ref{si-sec:si-order}). The order is retained only in a bounded range of write amplitudes. Below this range nothing is written. Above it each write brings the coordinates to a state that no longer depends on the state in which it found them, so each write erases what came before and only the last is recorded. Writes on uncoupled supports commute and leave no record of their order. States that depend on the order of applied fields were demonstrated in a zero-temperature Ising model by comparing histories that end in identical final fields~\cite{Croce2026}, which separates the order from the last field, and mechanical metamaterials show such states in experiment~\cite{Kwakernaak2023}. The ring shows the same with three writes in a substrate without frustration.

A record of order is therefore a dependence on the path taken through the protocol: two protocols with the same writes and the same final write reach different states. Material memory involves two different compositions. The composition of the transports of the couplings around the loops of the carrier decides which states exist (Sec.~\ref{sec:rules}), and the composition of the writes along the protocol decides which of these states is reached. The ring shows that the second does not require the first, since it records order without a frustrated loop.

\subsection{Relation between retention and rewriting times}
\label{sec:rewriting}

A bit is stored as the sign of a coordinate, and a write of bounded strength applies a force $F$ that favours the new sign. The retention time $T_h$ is the time after release at which the stored sign is still identified with probability $p^* = 0.9$. The rewriting time $T_w$ is the time under the opposite write after which the new sign is identified with the same probability.

\begin{figure}[!tbp]
\centering
\includegraphics[width=\linewidth]{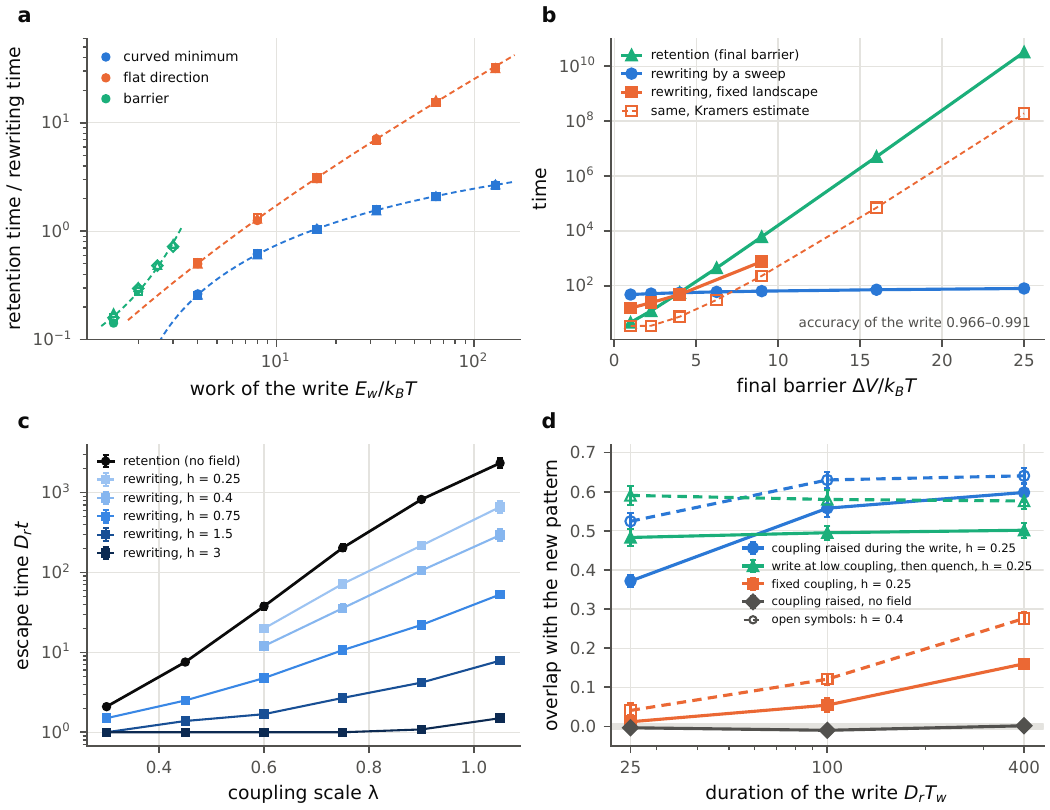}
\caption{\textbf{Retention and rewriting times, and writing in a changing landscape.}
\textbf{a}, Ratio of the retention time to the rewriting time against the work of the write, for a curved minimum ($\kappa = 0.5$, 1, 2: filled circles, open squares, open triangles), a flat direction ($s_0 = 1$, 2, 4) and double-well barriers ($\Delta V = 4$, 6, 8, $10\,k_BT$: circles, squares, triangles, diamonds). Dashed lines: Eq.~\eqref{eq:ratio-relaxation}, the prediction for a flat direction, and an exponential fit. The ratio depends only on the work of the write.
\textbf{b}, Writing at an instability: a pitchfork swept at rate 0.05 under a bias set for an accuracy of 0.99, against the final barrier. A complete rewrite by sweeping takes a fixed time, while the retention time at the final barrier grows exponentially. The same bias applied in a fixed landscape needs a time that grows with the barrier (filled symbols, simulated; open symbols, Kramers estimate, which lies below the simulations by factors of $\SweptKramersUnderRange$).
\textbf{c}, Capillary colloidal rotors ($N = 576$, five graphs): escape time from a stored angular state without a field (retention) and under a patterned field of uniform strength $h$ in units of $k_BT$ (rewriting), against the coupling scale. A strong field rewrites the state rapidly. Under weak fields the escape remains activated, and the ratio of the two times becomes independent of the coupling. Geometric means over graphs; error bars, standard errors.
\textbf{d}, Writing while the barriers are low, in the same network: overlap with the new pattern against the duration $T_w$ of the write (25, 100 and $400/D_r$), measured after a field-free interval of $200/D_r$ at $\lambda = 1.05$. The field was applied at $\lambda = 1.05$ (fixed coupling), while the coupling was raised linearly from 0.3 to 1.05 (coupling raised during the write), or at $\lambda = 0.3$ followed by an abrupt return to 1.05 (write at low coupling, then quench). The same increase of the coupling was also applied without a field. Filled symbols, $h = 0.25\,k_BT$; open symbols, $0.4\,k_BT$. Grey band, overlap between independent stored states. Means over five graphs; error bars, standard errors.}
\label{fig:rewriting}
\end{figure}

Without the write, the stored distinction is lost through spontaneous thermal fluctuations; under the write, the applied force biases those same fluctuations. In linear response, spontaneous decay and forced displacement obey the same dynamics~\cite{Onsager1931,Kubo1966}, and in an activated transition a weak force changes the escape rate through its work along the transition pathway~\cite{Kramers1940,Bell1978}. When the ratio $T_h/T_w$ is evaluated at the fixed identification threshold $p^*$, the timescale of the landscape cancels within each law of loss (SI Sec.~\ref{si-sec:si-rewriting}). In a harmonic minimum of curvature $\kappa$, the force shifts the stationary mean by $F/\kappa$, with characteristic energy $E_w = F^2/\kappa$, and
\begin{equation}
\frac{T_h}{T_w} = \frac{\ln\!\bigl(\sqrt{E_w/k_BT}\,/z^*\bigr)}{\ln\!\bigl[2/\bigl(1 - z^*\sqrt{k_BT/E_w}\bigr)\bigr]}, \qquad z^* = \Phi^{-1}(p^*),
\label{eq:ratio-relaxation}
\end{equation}
which grows only logarithmically with $E_w$; the curvature $\kappa$ drops out. Along a flat direction, a state written at $\pm s_0$ with work $E_w = F s_0$ gives $T_h/T_w \simeq E_w/(2 z^{*2} k_BT)$, linear in $E_w$. Behind an energy barrier, a weak force lowers the barrier in the favoured direction by $E_w \approx Fa/2$, where $a$ is the well separation, so that
\begin{equation}
\frac{T_h}{T_w} \propto e^{E_w/k_BT},
\label{eq:ratio-activation}
\end{equation}
which grows exponentially with $E_w$, while the barrier height $\Delta V$ itself cancels. One-coordinate Langevin simulations reproduce all three regimes (Fig.~\ref{fig:rewriting}a). Across barriers from $4$ to $10\,k_BT$, which lengthen retention $\LockLawActHoldFold$-fold, the ratio at fixed $E_w$ varies by at most $\LockLawActSpread\%$ and follows $e^{E_w/k_BT}$.

The times $T_h$ and $T_w$ are defined at fixed accuracy, and they differ from the mean first-passage times between basins. In a two-state reduction, $T_h$ and $T_w$ are linked to the mean escape times through the identification threshold $p^*$, the backward return flux and the equilibrium population bias. In the activated regime both kinds of time share the same leading barrier exponential; the prefactors that depend on the accuracy are kept in full in SI Sec.~\ref{si-sec:si-ratio-bound}.

For mean passage times, the variational theory of metastability gives an exact bound on any reversible carrier~\cite{Bovier2004Metastability}. Loss of the stored state corresponds to reaching a target region $\mathcal T$ from a disjoint stored region $\mathcal W$, and a static conservative write changes the potential from $V$ to $V - V_w$ at fixed temperature and mobility. Let $\overline T_h$ and $\overline T_w$ denote the mean first-passage times to $\mathcal T$, started from the normalized last-exit distribution on $\mathcal W$, of the unwritten and the written dynamics, respectively. Any reversible diffusion with finite passage integrals, and any finite reversible Markov network, then satisfies
\begin{equation}
\frac{\overline T_h}{\overline T_w} \le (1+\delta)e^{E_{\mathrm{eff}}/k_BT},\qquad
E_{\mathrm{eff}}=k_BT\ln\frac{\langle e^{V_w/k_BT}\rangle_{\mathrm{flux}}}{\langle e^{V_w/k_BT}\rangle_{\mathcal W}}.
\label{eq:ratio-bound}
\end{equation}
Here $\langle\cdot\rangle_{\mathcal W}$ is the Boltzmann average restricted to $\mathcal W$, and $\langle\cdot\rangle_{\mathrm{flux}}$ is the average weighted by the unwritten Dirichlet transition flux. To define these averages, let $\phi(x)$ be the unwritten committor, the probability of reaching $\mathcal W$ before $\mathcal T$, and let $\mathrm{d}\pi = e^{-V/k_BT}\mathrm{d}x$ be the unnormalized Boltzmann weight, so that the written weight is $\mathrm{d}\pi' = e^{V_w/k_BT}\mathrm{d}\pi$. The capacity is $\mathrm{cap}(\mathcal W,\mathcal T) = \mu k_BT\int|\nabla \phi|^2\,\mathrm{d}\pi$ for a diffusion with mobility $\mu$, or $\sum_{\{x,y\}} c_{xy}(\phi_x - \phi_y)^2$ over the undirected edges of a network with conductances $c_{xy} = \pi_x k_{xy}$. Potential theory expresses the mean exit time as~\cite{Bovier2004Metastability}
\begin{equation}
\overline T_h=\frac{\int \phi\,\mathrm d\pi}{\mathrm{cap}(\mathcal W,\mathcal T)},\qquad
1+\delta=\frac{\int \phi\,\mathrm d\pi}{\pi(\mathcal W)}.
\label{eq:mean-time}
\end{equation}
By the Dirichlet principle, $\mathrm{cap}'(\mathcal W,\mathcal T)$ minimizes the written quadratic form over all functions equal to one on $\mathcal W$ and zero on $\mathcal T$. Evaluating that form on the unwritten committor $\phi$ therefore bounds the written capacity from above, $\mathrm{cap}' \le \langle e^{V_w/k_BT}\rangle_{\mathrm{flux}}\,\mathrm{cap}$. The written numerator satisfies $\int \phi'\,\mathrm{d}\pi' \ge \pi'(\mathcal W) = \langle e^{V_w/k_BT}\rangle_{\mathcal W}\,\pi(\mathcal W)$. Dividing the two expressions proves Eq.~\eqref{eq:ratio-bound}. On a discrete network the bound applies with $c'_{xy} = c_{xy} e^{V_{w,xy}/k_BT}$, where $V_{w,xy}$ is the lowering of the transition-state energy along edge $\{x,y\}$ at fixed kinetic prefactors, and this lowering replaces $V_w$ inside $\langle\cdot\rangle_{\mathrm{flux}}$.

The prefactor $\delta \ge 0$ measures the equilibrium weight outside $\mathcal W$ that still commits back to $\mathcal W$ before reaching $\mathcal T$. It is small whenever $\mathcal W$ captures nearly the entire metastable basin. The exponent $E_{\mathrm{eff}}$ is an effective static energy bias between the stored basin and the unwritten transition region. It does not change when a constant is added to $V_w$, and it is bounded above by $\sup V_w - \inf V_w$. It is a static energy difference and differs from the work dissipated over a time-dependent cycle. In a weak-noise landscape dominated by a single stored minimum and a single transition saddle, $E_{\mathrm{eff}}$ reduces to the potential tilt between the minimum and the saddle. For a weakly tilted symmetric double well this recovers $Fa/2$, while the barrier height cancels.

Across \BoundLandscapes\ rough one- and two-dimensional potentials and random reversible networks, with mean passage times evaluated exactly, no ratio exceeded Eq.~\eqref{eq:ratio-bound}. The largest ratio reached \BoundMaxRatio\ of the bound, and barriers of $10\,k_BT$ or higher under weak tilts came within \BoundWeakWithin\% of it (SI Fig.~\ref{si-fig:si-bound}). The activated examples with a weak bias approach the exponential dependence of Eq.~\eqref{eq:ratio-bound}. The harmonic and flat-direction comparisons of Fig.~\ref{fig:rewriting}a concern instead the times defined at fixed accuracy, whose relation to mean passage times is given in SI Sec.~\ref{si-sec:si-ratio-bound}.

In a weakly tilted metastable memory dominated by a single escape path, raising the barrier lengthens the leading activated part of both retention and rewriting. Their ratio then depends on the bias along that path and on kinetic prefactors, and not on the bare barrier height alone. Magnetic recording meets this trade-off as the conflict between the thermal stability of the grains and the available writing field~\cite{Kryder2008HAMR}. Cascade models of synaptic memory meet it as the trade-off between plasticity and memory lifetime for bounded synaptic weights~\cite{Fusi2005Cascade,BennaFusi2016}. In a one-coordinate double well, this activated regime lies between two limits. A force strong enough to remove the barrier ($\Delta V \le Fa/\pi$ for a tilted sinusoidal potential) produces deterministic sliding. A barrier lowering $E_w$ below $\tfrac12\ln[p^*/(1-p^*)]\,k_BT$, for which the two wells differ by $2E_w$, cannot reach the identification accuracy $p^*$ even at equilibrium.

Equation~\eqref{eq:ratio-bound} limits the acceleration that a specified static bias supplies at fixed temperature and mobility. A sufficiently strong static bias can lower or remove the transition barrier and make rewriting rapid, while the state without the bias has a long lifetime. This case remains within the comparison of Eq.~\eqref{eq:ratio-bound}, at a large effective bias. When only weak writing fields are available, the two times can be separated in another way, by changing the landscape during writing.

Near the ridge of Sec.~\ref{sec:write}, a weak bias selects the basin before the retaining barrier has formed. The choice is then made with the accuracy $p^*$ within the detection time $2z^{*2}k_BT/(\mu F^2)$ of Sec.~\ref{sec:write}, which contains no barrier, whereas the same force acting behind a barrier $\Delta V$ needs a time that carries the factor $e^{(\Delta V-E_w)/k_BT}$. In the pitchfork sweeps of Fig.~\ref{fig:rewriting}b, the duration of a full cycle, an erasing sweep followed by a writing sweep, changes little over the tested final barriers. The retention time instead grows exponentially and exceeds the duration of the cycle above about $4\,k_BT$, and between 4 and $6\,k_BT$ the cycle becomes shorter than the rewriting time in the fixed landscape (SI Sec.~\ref{si-sec:si-rewriting}). Returning to an instability is one way of lowering a barrier during writing and restoring it afterwards. Heat-assisted magnetic recording lowers the anisotropy barrier by heating~\cite{Kryder2008HAMR}, and shape-memory polymers are programmed above their switching temperature and fixed by cooling.

Changing the mobility or the temperature during writing changes the corresponding transition rates. In the $8\,k_BT$ double-well examples, raising the mobility tenfold, or the temperature 1.5-fold, only during writing gives ratios that exceed the bound of Eq.~\eqref{eq:ratio-bound}, which assumes fixed temperature and mobility, by factors of \BoundMobilityExcess\ and \BoundTemperatureExcess, respectively (SI Fig.~\ref{si-fig:si-bound}). A static reduction of the barrier is included in $V_w$ in Eq.~\eqref{eq:ratio-bound}; a time-dependent sweep has its own passage dynamics. Photochromic switches provide a separate route: diarylethenes write through an excited electronic state and retain the ground-state isomer behind a thermal barrier~\cite{Irie2000Diarylethenes}. Section~\ref{sec:rewriting-carrier} develops modulation of the coupling for carriers in which the retaining rearrangements are collective.

\section{Predictions from structure}
\label{sec:rules}

When the transports of the couplings are composed around a closed cycle, the result decides whether every interaction can be satisfied at the same time. The symmetries under which the drift is equivariant fix the normal form of the local instability at which a state is written. When the couplings and the drive have compatible signs on every loop that the drive closes, return-point memory is exact. In directed regulatory networks, the signs of the cycles give necessary conditions for multistability and for sustained oscillation. Table~\ref{tab:rules} lists these eight structural rules. Exact state counts and retention lifetimes still require the full interaction energy, the noise intensity and the transition pathways (SI Sec.~\ref{si-sec:si-rules}).

\begin{table}[!tbp]
\centering
\footnotesize
\setlength{\tabcolsep}{4pt}
\renewcommand{\arraystretch}{1.15}
\begin{tabular}{>{\raggedright\arraybackslash}p{0.05\linewidth}>{\raggedright\arraybackslash}p{0.27\linewidth}>{\raggedright\arraybackslash}p{0.60\linewidth}}
\toprule
 & Structural input & Prediction \\
\midrule
R1 & Transports of the couplings around each loop & All couplings of a loop can be satisfied if and only if the composition of its transports has a fixed point \\
R2 & Gradient interactions, or signs of directed regulatory couplings & A frustrated energy cannot reach the minimum of every bond, and its stable compromises can be found only from the energy itself. In regulatory models, a positive circuit is necessary for isolated multistability and a negative circuit is necessary for sustained oscillation \\
R3 & A symmetry reversing a simple real critical mode & A nondegenerate odd reduced drift gives a pitchfork, and the sign of the cubic term decides its type. A degenerate critical space requires several amplitudes. A generic fold can be unfolded to a cusp when a suitable second parameter exists \\
R4 & A continuous symmetry acting nontrivially on equilibria & The orbit of equilibria has a neutral tangent. Noise projected along the tangent makes the state diffuse, and a compact orbit is eventually mixed. Writing and loss depend on the drive and the noise projected along the tangent \\
R5 & Reversible dynamics under a static energy bias, fixed temperature and mobility & The ratio of the mean times satisfies Eq.~\eqref{eq:ratio-bound}, with its prefactor and effective bias. Rescaling all mobilities by a common factor leaves the ratio unchanged \\
R6 & A transfer keeping the transport group, its action and reciprocity & The loop compatibility test keeps its form. The fluxes, stable states, noise and barriers must still be compared before the retained memory can be assigned to the same mechanism \\
R7 & A probe evolution and a specified hidden distinction & The distinction becomes measurable when the probe evolution maps it into the measured subspace: $\Pi U_{\mathrm{probe}}(t)\Pi^{\perp}\delta q\ne0$ in the linear description \\
R8 & Athermal quasistatic hysterons, with the drive included in the sign graph & If no loop through the drive is frustrated, a relabeling makes every coupling cooperative and the drive uniform, which gives no-passing and return-point memory. A frustrated loop through the drive removes the guarantee \\
\bottomrule
\end{tabular}
\caption{\textbf{Rules that predict a memory from structure.} Each rule names the structural input that it uses and the property of the memory that it predicts before any dynamics is solved. SI Table~\ref{si-tab:si-rules} gives the prior work for each rule and a diagnostic change of structure.}
\label{tab:rules}
\end{table}

\subsection{Loops of couplings}
\label{sec:loops}

When the minimum of a pairwise interaction defines an invertible map between the states of two neighbours, that map acts as a parallel transport across the bond. Ferromagnetic or nematic alignment transports a state unchanged, and antiferromagnetic exchange reverses the sign of a spin. The bond transports composed around a closed cycle give the loop holonomy of the cycle. Every bond of the cycle can reach its minimum at the same time if and only if the holonomy has a fixed state. For $\mathbb{Z}_2$ sign transports acting on nonzero spins, this requires the product of the bond signs around the cycle to be $+1$~\cite{Harary1953,Toulouse1977}. In directed gene-regulatory networks the signs of activation and repression have a different meaning. They record the sign of a derivative of the local Jacobian, not an invertible map between states, and they constrain the feedback loops through Thomas's rules, stated later in this subsection.

Couplings that depend only on the relative states of two elements can be frustrated, as antiferromagnetic exchange on a triangle is~\cite{Moessner2006}. Their frustration comes only from the signs of the couplings around a loop, and the positions of the elements cannot tune it continuously. For the positions to set the frustration, the coupling must be bond-directional: it must refer the orientations to the vectors that connect the elements, as compass couplings do in magnetic models~\cite{Nussinov2015Compass}. Anisotropic colloidal particles trapped at a fluid interface realize such a coupling. Their undulated contact lines deform the meniscus~\cite{Loudet2005,Stamou2000,Bowden1997}, and the resulting quadrupolar capillary interaction has the leading angular dependence~\cite{Stamou2000}
\begin{equation}
U_{ij} = -g \cos\bigl[2(\theta_i + \theta_j - 2\varphi_{ij})\bigr],
\label{eq:compass-coupling}
\end{equation}
where $\theta_i$ and $\theta_j$ are the in-plane orientations of the particles and $\varphi_{ij}$ is the orientation of the line that joins their centres. Nematic alignment depends on the difference $\theta_i-\theta_j$ and favours a common director. The capillary coupling depends on the sum and acts as a mirror: the orientation that one particle prefers is the reflection of the other particle's orientation across the line of centres. A bond is satisfied when $\Theta_i + \Theta_j = 2\Phi_{ij}$, with $\Theta = 2\theta$ and $\Phi = 2\varphi$. Summing the bond conditions around a closed loop of $n$ particles with alternating signs gives
\begin{equation}
\sum_{s=1}^n (-1)^{s+1}(\Theta_{i_s} + \Theta_{i_{s+1}}) = 2\sum_{s=1}^n (-1)^{s+1} \Phi_{i_s i_{s+1}}.
\label{eq:loop-sum}
\end{equation}
In an even loop each particle enters through its two bonds with opposite signs. The orientations then cancel identically, and a condition on the positions alone remains:
\begin{equation}
\Phi_{\gamma} = 4 \sum_{s=1}^n (-1)^{s+1} \varphi_{i_s i_{s+1}} \pmod{2\pi}.
\label{eq:loop-flux}
\end{equation}
The sign of $\Phi_{\gamma}$ depends on the orientation of the loop and on the choice of base point, but its vanishing, and with it the fixed-point condition, is gauge-invariant. Each capillary bond acts as a reflection, $\Theta_j = 2\Phi_{ij} - \Theta_i$, and reflections across distinct axes do not commute. An even number of reflections composes to a rotation by $\Phi_{\gamma}$. This rotation fixes no orientation unless $\Phi_{\gamma} = 0$, so a cycle with nonzero flux is geometrically frustrated, as in ice~\cite{Pauling1935}, frustrated magnets~\cite{Moessner2006} and artificial spin ice~\cite{Wang2006SpinIce}. An odd number of reflections composes to a net reflection, which always fixes the orientations along its mirror axis. An isolated cycle with an odd number of reflections can therefore be satisfied for any positions of the particles, while a cycle with an even number must meet the flux condition of Eq.~\eqref{eq:loop-flux}. The parity is the reverse of that of antiferromagnetic spin loops, in which odd loops are always frustrated and even loops never are. When several cycles share nodes in a network, the network is unfrustrated only if the holonomies of all its cycles, transported to a common base node, have a common fixed orientation.

On a frustrated loop, reciprocal couplings settle into a compromise configuration. For $N$ equal cosine bonds on a cycle with flux $\Phi_{\gamma} \in (-\pi, \pi]$, the energy is lowest when the angular mismatch is shared equally by the bonds. The ground-state energy then lies above the unfrustrated minimum by the fraction
\begin{equation}
f = 1 - \cos(\Phi_{\gamma}/N),
\label{eq:loop-frustration}
\end{equation}
exactly as in a ring of Josephson junctions threaded by a magnetic flux~\cite{TeitelJayaprakash1983}. Unequal bond strengths share the mismatch unevenly. Directed couplings are not reciprocal: in transcriptional networks one gene can regulate another without a symmetric back-action. Thomas's positive-feedback rule, proved by Soul\'e, states that when the drift has several nondegenerate stationary states, the interaction graph of the local Jacobian must contain a positive cycle somewhere in the domain~\cite{Thomas1981,Soule2003}. In a strongly connected regulatory system whose interaction signs are fixed, a stable periodic oscillation requires a negative feedback cycle that involves at least two variables~\cite{Snoussi1998}. Whether the attractors that these rules permit actually appear is decided by the regulatory gain.

The loop compatibility and feedback conditions were tested on \LoopCount\ isolated cycles and on \NetTotal\ random multi-loop networks of four to eight units (Fig.~\ref{fig:rules}a,b; SI Sec.~\ref{si-sec:si-rules}). The cycles were transcriptional rings of two to five genes with all combinations of repression signs, soft-spin rings of three to six elements with every number of antiferromagnetic bonds, and rotor rings of three to six particles with random positions and a mixture of alignment, anti-alignment and capillary reflection bonds. Every simulated cycle behaved as its structure predicted, and no simulated network contradicted its prediction. For gene networks the rules are necessary conditions, and networks for which they give no definite prediction are not counted. Positive gene cycles were bistable at the tested gain, and negative cycles oscillated or relaxed. Odd antiferromagnetic spin cycles left one bond unsatisfied. Rotor cycles with an even number of reflections followed Eq.~\eqref{eq:loop-frustration}, and those with an odd number of reflections remained unfrustrated whatever their geometry.

\subsection{Number of states and retention time}

Frustration changes how many compromise configurations a network has, while barrier heights and thermal noise decide how long a selected configuration lasts. In Brownian dynamics of the monolayer of capillary rotors, a written orientational pattern persists after the field is removed, while the global nematic order parameter stays near zero~\cite{BaulinColloidMemory}. In a binary reduction of the network, removing the loop frustration while keeping the interaction graph and the bond magnitudes fixed removes on average 5.58 states per graph that are stable against the reversal of any single element (95\% confidence interval 2.67--8.83), and it does not shift the retention lifetimes in a uniform direction~\cite{BaulinColloidMemory}. Loop incompatibility therefore multiplies the metastable basins among which a write can choose, and the escape times need a separate calculation of the barriers of the landscape.

\subsection{Symmetry and the kind of write}

As a control parameter is varied, a stable state can lose stability in only a few ways. When a single mode becomes critical, the drift reduced to that mode, $\dot u = a_0 + a_1 u + a_2 u^2 + a_3 u^3$, distinguishes them. If a symmetry of the drift reverses the critical mode, the reduced drift must be odd, and the state loses stability at a pitchfork~\cite{Golubitsky2002}. A weak bias then chooses between two equivalent states, and the write laws of Sec.~\ref{sec:write} apply with the bias and the noise projected on the critical mode.

A symmetry can also make the critical mode degenerate, so that two or more directions become critical at the same value of the control~\cite{Golubitsky2002}. This happens in a frustrated ring of an odd number of identical elements. The alternating pattern that the couplings favour does not close around the ring, and the two twisted modes nearest to it lose stability together. The drift then cannot be reduced to one coordinate. The write chooses one of several states that the symmetry relates within the critical eigenspace (when the symmetry is that of a square, the write chooses one of four states along the two symmetry axes of the critical plane, rather than one of two), and the write laws of Sec.~\ref{sec:write}, derived for one coordinate, do not apply as they stand. A difference between the elements splits the critical eigenvalues and restores a single critical mode. For a write it does so only when the splitting of the critical rates is larger than $\sqrt{r}$, where $r$ is the rate at which the sweep changes the critical growth rate. The choice is made over a time of order $r^{-1/2}$. A second direction whose relaxation time is comparable to this interval or longer stays dynamically active while the state is selected, so the write remains two-dimensional and does not follow the law of one mode (SI Sec.~\ref{si-sec:si-writes}).

Without such a symmetry nothing forbids a quadratic term, and the generic event is a fold. At a fold a state disappears and the system moves in one direction only, so the write is one-sided: it is a threshold. When a second parameter is accessible, it can bring a fold to a cusp, where the frozen reduced drift has no quadratic term. A symmetric write made at a finite rate must also account for the motion of the reference state during the sweep, as the chemical construction below makes explicit. A positive cubic term leaves the instability unsaturated, so the state jumps to a distant branch instead of choosing a nearby one. A complex pair of critical eigenvalues makes the state oscillate instead of choosing, as a negative loop requires.

A control that multiplies the whole drift changes the speed of the dynamics and none of its states, so the way the control parameter enters the drift is part of the structure as well. Such a control offers no write point, and the write must then be made by a field.

\subsection{Continuous symmetries and flat directions}
\label{sec:flat}

When a continuous symmetry acts nontrivially on an equilibrium state, it generates a continuous orbit of equilibria, and along the tangent of this orbit there is no deterministic restoring force. In a rotor network this happens whenever no cycle fixes a reflection axis. Rotating all rotors by a common angle, with signs that alternate across reflecting bonds, then leaves every bond energy unchanged (SI Sec.~\ref{si-sec:si-rules}). Thermal noise projected along this neutral tangent makes a written phase or angle diffuse. On a compact orbit the long-time mixing is governed by its non-zero eigenvalues, and the unwrapped angle spreads diffusively, so that the ratio of retention to rewriting time grows linearly with the work of the write (Sec.~\ref{sec:rewriting}). Time-translation invariance along an autonomous limit cycle produces the same neutral diffusion of the phase in time (Sec.~\ref{sec:carriers}).

\begin{figure}[!tbp]
\centering
\includegraphics[width=\linewidth]{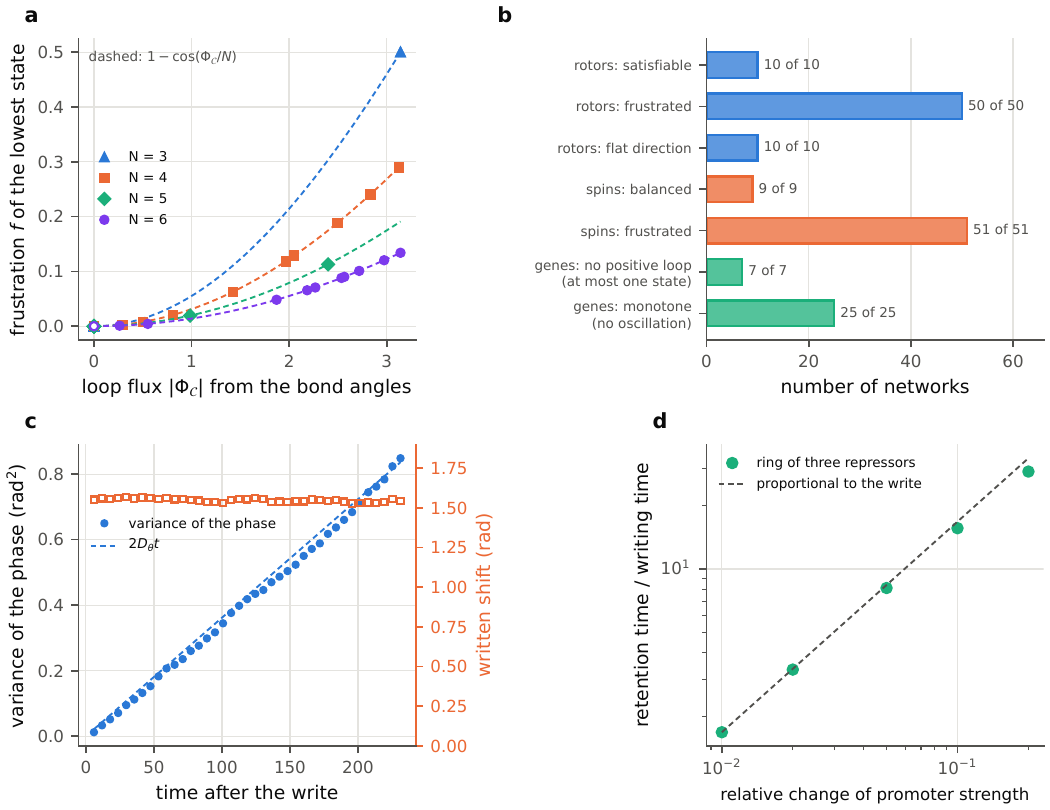}
\caption{\textbf{Predictions from structure.}
\textbf{a}, Frustration of rotor loops of three to six bonds, with random positions and bond types, against the loop flux computed from the bond angles. Dashed line: Eq.~\eqref{eq:loop-frustration}. Open symbols at zero flux: loops with an odd number of reflections.
\textbf{b}, Random networks of four to eight rotors, spins or genes (\NetPerKind\ of each kind). Outlined bars: the number of networks predicted in each class; filled bars and numbers: the networks for which the simulation agrees with the prediction. The class of rotor networks with a flat direction overlaps the other two rotor classes. For genes the rules are necessary conditions; the other 28 gene networks have no definite prediction and are not shown.
\textbf{c}, Phase of the ring of three repressors after a written shift: variance of the phase (circles, left axis; dashed line, $2D_\theta t$) and the shift itself (squares, right axis), against time.
\textbf{d}, Ratio of the retention time of a phase shift to the time needed to write it by a change of the promoter strength, against the relative change. Dashed line: proportionality.}
\label{fig:rules}
\end{figure}

\subsection{Loops through the drive and the return to a turning point}
\label{sec:return-point}

Many of the memories of Sec.~\ref{sec:introduction} record the turning points of a scalar drive. Their elements switch at thresholds of the drive and sit behind barriers large compared with the noise, so the state depends on the sequence of turning points and not on the rate of driving (Sec.~\ref{sec:construction}). The exact form of this record is return-point memory. When the drive turns at $H_1$, makes an excursion that does not pass $H_1$ and comes back to it, the material returns exactly to the state that it had at $H_1$. The excursion is erased, and the turning points that enclose it are retained~\cite{Sethna1993Hysteresis,Mayergoyz1991,Keim2019Memory}. Repeated excursions nest. The record is an alternating sequence of maxima and minima of the drive, each pair enclosed by the one before. A drive that passes an earlier turning point erases it together with every turning point recorded after it, which is the wiping-out property of hysteresis~\cite{Mayergoyz1991}. Independent elements, as in the Preisach model, have this property by construction. For interacting elements it follows from the no-passing property~\cite{Middleton1992,Sethna1993Hysteresis}. Suppose that every coupling favours equal states of the elements that it joins, and that a larger drive favours the upper state of every element. A configuration that lies above another at every element then stays above it under any common history of the drive. The dynamics preserves this ordering of configurations, and the return to a turning point follows from it. Couplings of both signs remove the guarantee, and most of the transition graphs of three interacting elements violate return-point memory~\cite{vanHecke2021Hysterons}.

Counting the drive as an element of the network turns this condition into a statement about loops. Let the drive act on element $i$ with the sign $\eta_i$, positive when raising the drive favours the upper state of $i$, and let $J_{ij}$ be the sign of the coupling between $i$ and $j$. In the network that contains the drive as an element joined to every other element, each bond $ij$ closes a loop through the drive with the product of signs $\eta_iJ_{ij}\eta_j$. If no loop of this network is frustrated, relabeling the states as $\sigma_i = \eta_is_i$ makes every coupling positive and the drive uniform. The relabeled elements obey the no-passing property, and the return to a turning point is exact, whatever the order in which the elements of an avalanche switch (SI Sec.~\ref{si-sec:si-return}). The condition is rule R1 applied to the realization extended by its protocol (R8 in Table~\ref{tab:rules}), and it is the condition under which a system driven by an input is monotone~\cite{AngeliSontag2003}. Because every element is driven, a frustrated loop of couplings always contains a bond whose loop through the drive is frustrated. The converse is false: a network whose loops of couplings are all satisfied can still have frustrated loops through the drive.

\begin{figure}[!tbp]
\centering
\includegraphics[width=\linewidth]{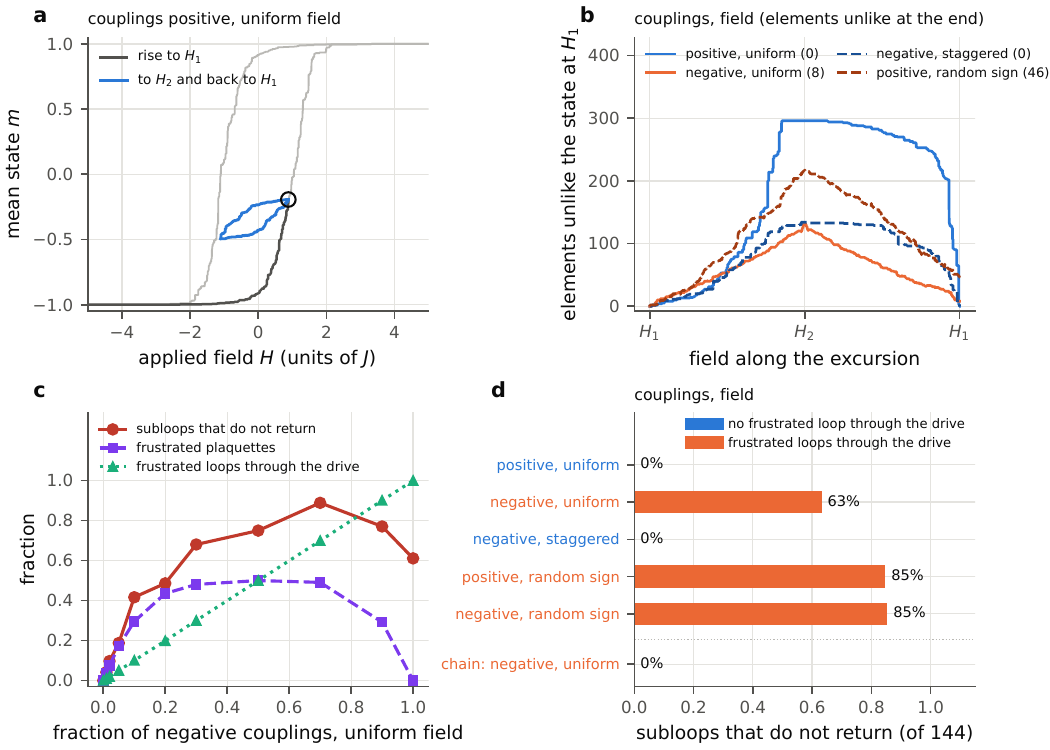}
\caption{\textbf{Return to a turning point of the drive.} Zero-temperature random-field Ising model on a periodic square lattice of $\RpmN$ elements, with couplings of unit strength and random fields of width $\RpmR$, driven quasi-statically.
\textbf{a}, Mean state against the drive for positive couplings in a uniform field: the major loop (grey), the rise from a large negative field to $H_1$ (dark), and an excursion to $H_2$ and back (blue), which ends in the state at $H_1$ (circle).
\textbf{b}, Number of elements whose state differs from their state at $H_1$ along one excursion from $H_1$ to $H_2$ and back, for four combinations of the signs of the couplings and of the drive; the number at the return is given in parentheses. Blue and dashed dark blue: no frustrated loop through the drive. Orange and dashed red: frustrated loops through the drive.
\textbf{c}, Fraction of subloops that do not return, against the fraction of negative couplings in a uniform field, shown with the fractions of frustrated plaquettes and of frustrated loops through the drive and one bond.
\textbf{d}, Fraction of subloops that do not return for five combinations of couplings and drive on the lattice, and for an antiferromagnetic chain of $\RpmChainL$ elements in a uniform field (below the dotted line), in which every loop through the drive is frustrated. Blue: no frustrated loop through the drive; orange: frustrated loops through the drive. Each bar counts $\RpmSubloops$ subloops.}
\label{fig:return-point}
\end{figure}

Simulations of the zero-temperature random-field Ising model~\cite{Sethna1993Hysteresis}, in which threshold elements with random thresholds are coupled on a square lattice, separate the two conditions (Fig.~\ref{fig:return-point}; SI Sec.~\ref{si-sec:si-return}). With positive couplings in a uniform field every subloop returns exactly. So does the antiferromagnet in a staggered field, which is the same network relabeled. The antiferromagnet in a uniform field frustrates no plaquette, because the lattice is bipartite, but it frustrates every loop through the drive, and most of its subloops fail to return. Positive couplings under a drive of random sign fail as well, and a small fraction of negative couplings among positive ones already breaks the return of some subloops. Whether interacting hysteretic elements return to their turning points is therefore decided by the loops that close through the drive, and not by the frustration of the couplings alone. A frustrated loop through the drive permits a failure but does not force one. Random antiferromagnetic chains started from a large field return exactly~\cite{Deutsch2004ReturnPoint}, and the same dynamics reproduces this result. The absence of frustration in the extended network is sufficient for return-point memory, but it is not necessary.

\section{Carriers}
\label{sec:carriers}

\subsection{Arrays of in-plane dipoles}

Point dipoles confined to rotate in a plane provide a second carrier with a bond-directional coupling. With the doubled angle $\psi = 2\theta$, the rotor energy becomes $-\sum_{ij}[J w^{a}_{ij}\cos(\psi_i - \psi_j) + g w^{c}_{ij}\cos(\psi_i + \psi_j - 4\varphi_{ij})]$, where $J$ and $g$ are the strengths of the alignment and the bond-directional coupling, $(J,g)=\lambda\,(4,5)\,k_BT$ for the colloidal rotors (Methods), and $w^{a}_{ij}$ and $w^{c}_{ij}$ are the weights of the two terms on the bond $ij$. Two in-plane point dipoles with moments along $\psi_i$ and $\psi_j$ interact as
\begin{equation}
\begin{split}
E_{ij} &= \frac{m^2}{r_{ij}^3}\Bigl[\cos(\psi_i - \psi_j) - 3\cos(\psi_i - \varphi_{ij})\cos(\psi_j - \varphi_{ij})\Bigr] \\
&= -\frac{m^2}{r_{ij}^3}\Bigl[\tfrac12\cos(\psi_i - \psi_j) + \tfrac32\cos(\psi_i + \psi_j - 2\varphi_{ij})\Bigr].
\end{split}
\label{eq:dipolar-map}
\end{equation}
The second line has the form of the rotor energy. Each pair interaction is the sum of two terms. The first, which depends on $\psi_i - \psi_j$, aligns the two moments as in a ferromagnet. The second, which depends on $\psi_i + \psi_j - 2\varphi_{ij}$, is bond-directional: it reflects one moment across the axis joining the two particles. Both terms decay with distance as $(r_0/r)^3$, and their ratio is fixed at $J : g = 1 : 3$. An array of in-plane dipoles is thus an exact rotor network of the same interaction family. It is not, however, an exact dynamical copy of the capillary monolayer when the dipoles are placed at the colloid positions. Under $\psi = 2\theta$ such a copy would require the doubled bond angles $2\varphi_{ij}$, and planar positions cannot supply doubled angles around a closed cycle. The transfer changes three things instead. The coupling ratio changes from $4:5$ to $1:3$. The interaction range extends to $r^{-3}$ (truncated at 3.5 spacings in the simulations; Methods), with coupling scale $\lambda_d = m^2/(r_0^3 k_BT)$. At the same particle positions, the loop flux is halved to $\Phi_{\gamma}^{\mathrm{dip}} = 2\sum_s (-1)^{s+1}\varphi_{i_s i_{s+1}}$. Two properties survive the transfer: the bond-directional term still transports an orientation by a reflection, and the single-loop compatibility test (rule R6) keeps its parity for the bond-directional term: an isolated odd cycle of that term can still satisfy all its bonds, while an even cycle is frustrated by the particle positions alone. The dipolar energy also contains the alignment term. With both terms the minimum of a single bond places the two moments along the bond, so a loop whose bonds are not collinear, an odd loop included, generally cannot reach the minimum of every bond. Whether the compromise basins remain invisible in the macroscopic magnetization, and whether they retain written states by activation, has to be checked on the dipole energy landscape itself.

Brownian simulations of point dipoles at the colloid positions show the properties that can be measured (Fig.~\ref{fig:transfer}a; SI Sec.~\ref{si-sec:si-dipoles}). Equilibrated states carry no net magnetization at any coupling. The net moment is $\DipSRange$, close to the $1/\sqrt{N}$ floor of random orientations. The orientational order measured relative to the bonds grows with $\lambda_d$. The stored pattern is therefore invisible in the global dipole moment, in the same way as the colloidal pattern is invisible in the nematic director. The retention lifetime grows exponentially with $\lambda_d$. Weak fields rewrite stored states through the same activated channel, and as the field weakens the logarithmic slope of the rewriting time approaches the retention slope.

\subsection{Write points}
\label{sec:write-points}

A write through an instability needs a point at which a stable state loses stability, or ceases to exist, as a control parameter changes. If a single direction becomes critical at that point, the drift reduced to this direction can be compared with the pitchfork of Sec.~\ref{sec:write}, the symmetric write. The terms by which the reduced drift differs from the pitchfork are the obstruction to a symmetric write. They are interpreted as in Sec.~\ref{sec:rules}. A quadratic term or a bias makes the write one-sided. A positive cubic term makes the state jump to a distant branch. A complex pair of eigenvalues makes the state oscillate. When two or more directions become critical together, in a degenerate critical mode, the drift cannot be reduced to one coordinate. When an accessible second parameter leads to a cusp, the quadratic term vanishes in the frozen normal form. A sweep at finite rate moves the reference state, and this motion adds a drift; an unbiased sweep must compensate it. Table~\ref{tab:write-points} lists the results, and Methods describes how the write points were located.

\begin{table}[!tbp]
\centering
\footnotesize
\setlength{\tabcolsep}{3pt}
\renewcommand{\arraystretch}{1.15}
\begin{tabular}{>{\raggedright\arraybackslash}p{0.2\linewidth}>{\raggedright\arraybackslash}p{0.16\linewidth}>{\raggedright\arraybackslash}p{0.2\linewidth}>{\raggedright\arraybackslash}p{0.19\linewidth}>{\raggedright\arraybackslash}p{0.18\linewidth}}
\toprule
Realization & Write point & Kind & Obstruction & Removed by \\
\midrule
Pitchfork normal form & $\epsilon = 0$ & Supercritical pitchfork & None & -- \\
Genetic toggle~\cite{Gardner2000Toggle} & $\alpha = \ToggleAlphaC$ & Supercritical pitchfork & None & -- \\
Four repressors in a ring & $\alpha = \RingFourAlpha$ & Supercritical pitchfork & None & -- \\
Three repressors in a ring~\cite{Elowitz2000Repressilator} & $\alpha = \RingThreeAlpha$ & Hopf & Rotation (negative loop) & Changing the sign of one bond \\
Schl\"ogl reactor~\cite{Schlogl1972} & $b = \SchloglFoldLow$, $\SchloglFoldHigh$ & Folds & Quadratic term; bias & Cusp at $(a, b) = (\SchloglCuspA, \SchloglCuspB)$ \\
Two adaptive tubes, equal lengths~\cite{Tero2007Physarum} & $n_{\mathrm H} = \TubesMuC$ & Subcritical pitchfork & Positive cubic term & A saturating term \\
Two adaptive tubes, unequal lengths & $n_{\mathrm H} = \TubesUnequalMu$ & Fold & Bias; quadratic term & Equal lengths \\
Capillary rotors & Field at which the stored state is lost & Loss of stability under the field & Write bounded by the retaining landscape & Raising the coupling during the write \\
Three compartments & None & Relaxation to one state & -- & -- \\
\bottomrule
\end{tabular}
\caption{\textbf{Write points of realizations on different carriers.} Each row gives the point at which a stable state loses stability or ceases to exist along the control parameter of the realization; the kind of instability, obtained from the drift reduced to the critical direction; the obstruction to the pitchfork of Sec.~\ref{sec:write}; and the change that removes the obstruction. For the rotors the control is the coupling scale. It multiplies the whole drift and so changes no state. The write point of the rotors is the field at which the stored state disappears.}
\label{tab:write-points}
\end{table}

A genetic toggle consists of two genes that repress each other~\cite{Gardner2000Toggle}, and it realizes the pitchfork without obstruction. In its symmetric state the two genes are expressed equally. This state becomes unstable along the antisymmetric direction, in which one gene rises as the other falls. Exchanging the two genes reverses this direction and leaves the dynamics unchanged, so the reduced drift has no quadratic term. The swept-write law of Eq.~\eqref{eq:sweep-law} then carries over to the toggle. The bias and the noise enter as their projections on the critical direction, and the sweep rate enters as the rate at which the sweep changes the linear coefficient, taken along the moving symmetric state. Simulated toggles follow the transferred law to within $\ToggleLawMaxDev$ for predicted accuracies from 0.6 to 0.95 (Fig.~\ref{fig:transfer}c). The law fails only when the noise becomes comparable to the growth of the barrier behind the choice. The state can then cross back after it has chosen (SI Sec.~\ref{si-sec:si-writes}).

Schl\"ogl's model~\cite{Schlogl1972}, $\dot x=-x^3+ax^2-k_3x+b$, has two stable concentrations, and crossing a fold in the feed $b$ selects one of them. Following a fold while the autocatalytic feed $a$ changes leads to a cusp, where the frozen cubic can be centred on $x=a/3$ (Fig.~\ref{fig:transfer}b). This centre moves during a sweep. In the shifted coordinate $y=x-a(t)/3$ the velocity of the centre contributes $-\dot a/3$ to $\dot y$. A sweep along the tangent that removes the frozen bias is therefore not enough to give an unbiased write in time.

The following feed compensates the motion of the centre:
\begin{equation}
\begin{aligned}
a(t)&=\sqrt{3(k_3+rt)},\\
b(t)&=\frac{k_3a(t)}{3}-\frac{2a(t)^3}{27}
       +\frac{\dot a(t)}{3}+h .
\end{aligned}
\label{eq:schlogl-write-feed}
\end{equation}
In a neighbourhood of the cusp where the feed is positive, this feed gives exactly $\dot y=rt\,y-y^3+h$. The translation leaves additive concentration noise unchanged. The swept-write law of Eq.~\eqref{eq:sweep-law} therefore applies to this compensated local model, under the same weak-noise conditions as for the pitchfork. SI Sec.~\ref{si-sec:si-writes} derives the frozen feed curve and the correction that depends on the rate.

Adaptive transport networks store memory in the conductances of their tubes. In the two-tube model of \emph{Physarum}~\cite{Tero2007Physarum}, the symmetric state, with equal flow in both tubes, loses stability through a subcritical pitchfork with a positive cubic coefficient, and one tube captures the flow. Unequal tube lengths break the exchange symmetry. The unbiased bifurcation then unfolds into a one-sided fold threshold that favours the shorter tube. A symmetric write without bias needs equal lengths. Retaining either path once it has formed does not. With flux reinforcement $f(Q) = Q^{n_{\mathrm H}}/(1 + Q^{n_{\mathrm H}})$ and $n_{\mathrm H} > 1$, both states in which a single tube carries the flow remain locally stable for any finite positive tube lengths. A difference in length shifts the basin boundaries and the write thresholds, but it does not remove the unfavoured state (SI Sec.~\ref{si-sec:si-writes}). Rings of transcriptional repressors show the feedback loop rule at a write point. Three repressors form a negative cycle, and its symmetric state loses stability through a complex conjugate pair of eigenvalues into a limit cycle~\cite{Elowitz2000Repressilator}. Four repressors form a positive cycle, which bifurcates along a real mode into two alternating expression patterns. The three-compartment exchange of Sec.~\ref{sec:reading} has no write instability at all. It relaxes to a unique equilibrium and stores its history only transiently, in its projection kernel. Taken together, these examples show that the existence of retained states is separate from the symmetry and bifurcation structure through which a protocol selects one of them~\cite{CodiscoveryCompanion}.

Predictor--corrector continuation and centre-manifold reduction extend this analysis to a survey of 110 dynamical models. The survey contains 79 systems of ordinary differential equations compiled from arXiv sources~\cite{HyperionCompanion} and 31 semiclassical models of quantum-computing and switching hardware~\cite{FieldBridge}. The 79 systems are those of 103 compiled specifications that carry the parameter values of their source and are distinct: 13 specifications without source values and 11 that repeat a system already present were left out. Of the 110 models, 101 declare a control parameter, and 44 of them cross at least one instability within the scanned interval. These crossings give 78 critical points. Saddle-node folds occur in 17 models, supercritical pitchforks in nine, Hopf pairs in six, neutral lines of equilibria in three, and critical eigenspaces made degenerate by symmetry in five; a model can show several kinds, and subcritical pitchforks, transcritical crossings and points where the quadratic and cubic terms vanish together make up the rest (SI Sec.~\ref{si-sec:si-published}). The 31 hardware models were chosen because they describe switching and measurement elements.

The 31 hardware models were followed beyond the linear instability: through the nonlinear separation of the branches and through an interval after the write in which the control is kept at its final value (SI Sec.~\ref{si-sec:si-published}). In these 31 models, calculations of the stochastic passage that follow all slow modes across the write window classify 41 write outcomes at the critical points. In 29 outcomes the branch is selected during the stage of linear amplification, in quantitative agreement with the reduced one-mode law Eq.~\eqref{eq:sweep-law}. For seven of them, the driven Kerr resonators and a Rydberg ensemble, the agreement requires the shift that the fluctuations about the path of the sweep give the slow coordinate at second order. In five the branch selection is completed during the nonlinear stage, and the written state is retained afterwards. Four of these five lie at points where two directions are critical together. In six the state keeps making transitions between the wells after the sweep, with the control kept at its final value, because the wells are shallow and their occupation approaches equilibrium. In one the state has not settled at the end of the sweep. This fourth regime occurs in a bistable optomechanical resonator. At threshold the resonator has a single static critical eigenvalue, but three of its directions relax slowly on the time scale of the sweep. These slow directions delay the commitment to a branch, without any thermal switching between wells that have already formed.

A fold can be unfolded to a codimension-two cusp, so that the local cubic drift has no quadratic term at threshold. Even then, the reference branch moves during a finite sweep, and this motion can introduce an effective offset bias. Several slow modes within $\sqrt{r}$ of the critical growth rate rotate the state and redistribute probability before saturation. Shallow barriers after the sweep permit thermal crossing back. Of the nine nonlinear-stage and equilibrium outcomes at which the leading one-mode law fails, local deterministic corrections to the reduced normal form are verified at two; across the full nonlinear separation of the branches, the coupled slow manifold must be kept~\cite{FieldBridge}.

\subsection{Selective erasure through two nearby instabilities}
\label{sec:selective-erasure}

A weak bias breaks the symmetry between two mirror-image patterns, so they lose stability at slightly different couplings. A temporary increase of the coupling to a value between the two thresholds therefore removes one pattern and spares the other. Consider three overdamped amplitudes coupled by diffusion,
\begin{equation}
\mathrm dq_i=\left[(\varrho-g_i)q_i-q_i^3+
\chi\sum_{j\ne i}(q_j-q_i)+h_i\right]\mathrm dt+\sqrt{2D}\,\mathrm dB_i ,
\label{eq:selective-model}
\end{equation}
with site asymmetries $g = (0, 0.02, 0.25)$, a baseline diffusive coupling $\chi = 0.01$ and a weak bias on the first site, $h \approx (1.58\times10^{-3}, 0, 0)$. Sweeping the control parameter to $\varrho = 0.5$ and then removing the bias leaves a fixed final landscape with eight verified stable equilibria. A temporary pulse in the coupling $\chi$ can be applied at an intermediate control value, with $\chi$ returned to $0.01$ before the sweep finishes. The pulse leaves the final equations of motion unchanged, yet it changes which stored pattern survives.

In a single realization, let $m = (q_1 + q_2 + q_3)/3$ be the mean amplitude and $S = \tfrac12\sum_i(q_i - m)^2$ the dispersion of the three amplitudes about it. Raising the coupling by $\delta\chi$ contributes
\begin{equation}
\left.\dot m\right|_{\mathrm{pulse}}=0,\qquad
\left.\dot S\right|_{\mathrm{pulse}}=-6\,\delta\chi S .
\label{eq:selective-mixing}
\end{equation}
Diffusive mixing exerts no direct force on the mean $m$. The cubic drift of $m$, however, depends on the differences between the sites. The mixing generator and the unperturbed nonlinear drift therefore do not commute on $m$, and this coupling makes the fate of the stored pattern depend on the timing and amplitude of the mixing pulse.

The patterns are labelled by the signs of the first two amplitudes $(q_1, q_2)$ in the equilibrium catalogue that the three modes share. At the intermediate control value $\varrho = 0.16$, continuation in $\chi$ shows that the two mixed-sign patterns $(+,-)$ and $(-,+)$ lose stability at two distinct nondegenerate saddle-node folds, $\chi_- \approx 0.0280$ and $\chi_+ \approx 0.0293$. The small bias $h_1$ separates the two fold thresholds. With $h = 0$ they coincide, and reversing the sign of $h_1$ interchanges which pattern disappears first. A pulse amplitude at the midpoint of the calculated interval, $\chi \approx 0.0286$, destabilizes the targeted $(+,-)$ branch, while the protected $(-,+)$ branch stays inside its stable basin.

The erasure protocol keeps the control at $\varrho = 0.16$ and $\chi \approx 0.0286$ for $483$ time units, then restores $\chi = 0.01$ and completes the sweep to $\varrho = 0.5$. This interval is twice the local deterministic passage time through the bottleneck left by the vanished $(+,-)$ fold. In Langevin integrations at low noise ($D = 1.07\times 10^{-7}$), all 256 reference trajectories that start in the targeted $(+,-)$ pattern are rewritten to the uniform $(+,+)$ state, whereas all 256 trajectories that start in the protected $(-,+)$ pattern remain in $(-,+)$. Matched control protocols show that both parts of the calculation are needed. A weaker pulse, below $\chi_-$, preserves both patterns. A stronger pulse, above $\chi_+$, erases both. A shorter pulse at the midpoint ends before the trajectories have passed the bottleneck of the fold (Methods; SI Sec.~\ref{si-sec:si-selective}).

\subsection{Phase of a limit cycle}

When the only attractor of a system is a stable limit cycle, there is no second state in which a write can be stored. The system still retains the time at which a perturbation arrived. The equations of an autonomous oscillator do not depend explicitly on time, so every time-shifted copy of the periodic solution is again a solution. The phase along the cycle is therefore a direction without a restoring force: nothing pulls a shifted oscillator back to its former phase. A shift of the phase $\theta$ is retained, and lost, by phase diffusion, $\langle\delta\theta^2\rangle = 2D_\theta t$. This is diffusion on a compact carrier, and the loss becomes exponential once the spread of the phase reaches about one radian, where the slowest Fourier mode on the circle, $e^{-D_\theta t}$, takes over. In the ring of three repressors of Table~\ref{tab:write-points}, a written shift of $\pi/2$ was still $\PhaseSepEnd$~rad after $\PhasePeriods$ periods, and the variance of the phase grew linearly in time (Fig.~\ref{fig:rules}c; SI Sec.~\ref{si-sec:si-phase}). A phase can also be written by a temporary change of the frequency. The shift then accumulates at a rate proportional to the change of frequency, whereas the retention time of the shift does not depend on the write. The ratio of retention time to write time therefore grows linearly with the strength of the write, as Sec.~\ref{sec:rewriting} gives for a flat direction (measured log--log slope $\PhaseLockSlope$; Fig.~\ref{fig:rules}d). A sustained oscillation requires a drive, because under detailed balance a system relaxes to equilibrium. The precision of the phase is limited by the free energy dissipated per period~\cite{Cao2015Oscillations,BaratoSeifert2016Clocks}.

The flat direction acquires a restoring force when the oscillator is coupled to an amount $x$ that shifts its frequency and whose growth it gates, $\dot x = x(a - bx + g\cos\varphi)$, $\dot\varphi = \omega - hx$, with $\varphi$ the phase relative to an external rhythm. Linearization about the entrained state $(x^*,\varphi^*)$ gives $\lambda^2 + bx^*\lambda + ghx^*|\sin\varphi^*| = 0$ for a small deviation: the restoring coefficient of the phase and its damping are both proportional to the amount, and at small amounts a deviation decays at the rate $bx^*/2$. While the amount is small the phase is retained as on a free cycle, and the phase present at that stage decides whether the amount grows or vanishes; as near the ridge of Sec.~\ref{sec:write}, the selection is made before the structure that retains its outcome has formed.

\subsection{Further carriers}

Anisotropic inclusions in a lipid bilayer impose saddle-like curvature deformations on the membrane. Through the elasticity of the membrane they interact with the same quadrupolar, bond-directional angular dependence as capillary colloids at a fluid interface~\cite{DommersnesFournier1999}. The bending rigidity and tension of the membrane and the viscosity of the bilayer set the energy scale and the rotational diffusion. Meshing gears and hinged rotating linkages are another carrier. They enforce $\theta_j = c_{ij} - \theta_i$, so they transport a rotation by a reflection across an offset angle $c_{ij}$ fixed by the geometry of the teeth or hinges. Composing these reflections around a closed loop gives the same parity rule. An isolated odd cycle fixes a reflection axis and becomes rigid. An even cycle accumulates a rotation holonomy equal to the alternating sum of the offsets. It turns freely when this sum vanishes and jams when it does not~\cite{Meeussen2016Geared}. When the linkages are elastically compliant, an incompatible offset flux strains the even cycle into compromise configurations. The discrete metastable minima of these configurations depend on the nonlinear elastic energy.

\begin{figure}[!tbp]
\centering
\includegraphics[width=\linewidth]{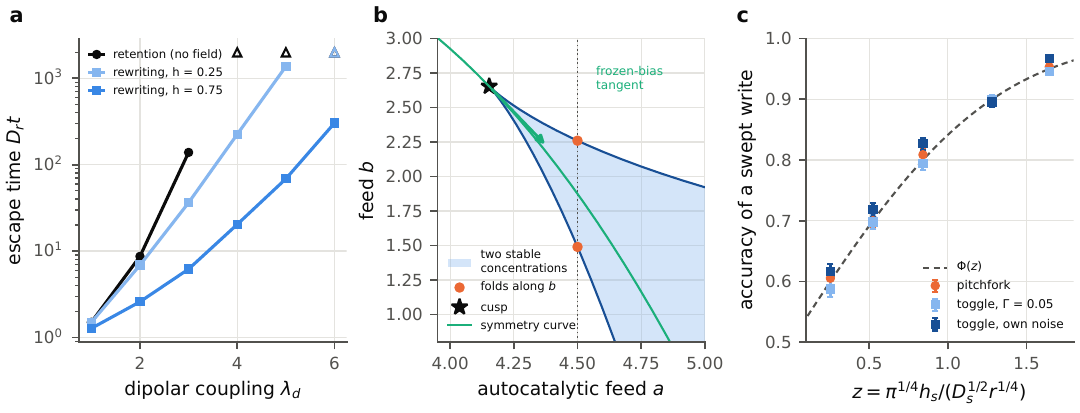}
\caption{\textbf{The colloidal memory and its write on other carriers.}
\textbf{a}, Arrays of in-plane point dipoles at the positions of the capillary rotors (five graphs). Escape time from a stored state, in units of the rotor diffusion time, against the dipolar coupling: without a field (retention) and under fields of 0.25 and $0.75\,k_BT$ (rewriting). Open triangles mark couplings at which at least one graph retained the state for the whole recording. The net moment is $\DipSRange$ at every coupling, close to the $1/\sqrt{N}$ of random orientations. Values are geometric means over the graphs.
\textbf{b}, Schl\"ogl chemistry in the plane of the two feeds. Shaded: the region with two stable concentrations. Circles: the folds crossed along $b$ at $a=4.5$. Star: the cusp reached by following a fold. The green curve centres the frozen cubic, and the arrow is its tangent at the cusp. Eq.~\eqref{eq:schlogl-write-feed} gives the compensated writing feed at finite rate.
\textbf{c}, Accuracy of a swept write in the genetic toggle and in the pitchfork, against the argument of the transferred law, $z = \pi^{1/4}h_s/(D_s^{1/2}r^{1/4})$, where the subscript $s$ denotes the projection on the critical direction $u$, at weak noise ($\Gamma = 0.05$) and at the noise of the toggle itself ($\Gamma = \ToggleOwnGamma$). Dashed: $\Phi(z)$. Error bars are binomial standard errors.}
\label{fig:transfer}
\end{figure}

\section{Rewriting on a given carrier}
\label{sec:rewriting-carrier}

The one-coordinate models of Sec.~\ref{sec:rewriting} escape over a single saddle. In a many-body network a stored pattern escapes, and is rewritten, by collective rearrangements in which several particles move together. The capillary rotor monolayer shows how the retention time and the rewriting time scale with the interaction strength when the barriers are collective. Point-dipole arrays show whether the same relation carries over to a second substrate.

In the capillary rotor monolayer every interaction energy is proportional to the dimensionless coupling $\lambda$, and a stored orientational pattern decays by activated collective escapes~\cite{BaulinColloidMemory}. To rewrite, an equilibrated parent pattern was driven toward an independently equilibrated target pattern by a site-patterned field of uniform amplitude $h$, aligned at each site with the target angle (Methods). Without a field, the logarithm of the time for the overlap to escape grows linearly with $\lambda$. This is the signature of activated retention across barriers proportional to $\lambda$. Under the rewriting field the network shows both regimes of Sec.~\ref{sec:rewriting} (Fig.~\ref{fig:rewriting}c). A strong field ($h = 3\,k_BT$ per rotor) removes the barriers and rewrites the pattern within $\ColloidStrongRewriteRange\,D_r^{-1}$ up to $\lambda = \ColloidStrongLambdaMax$. Weak fields ($h = 0.25$ and $0.4\,k_BT$) leave the escape activated, so retention and rewriting slow down together as $\lambda$ increases. At the strongest coupling, the logarithm of the ratio of retention time to rewriting time increases by $\ColloidWorkPerField$ per $k_BT$ of applied field. In a weak-bias activated description (Eq.~\eqref{eq:ratio-activation}) this slope is the displacement of a collective transition in which roughly five rotors reorient together (SI Sec.~\ref{si-sec:si-colloids}).

Because every interaction barrier scales with $\lambda$, lowering $\lambda$ softens all collective barriers at once, as heating lowers the anisotropy barriers of a magnet. A weak patterned field ($h = 0.25$ or $0.4\,k_BT$) applied at the retention coupling $\lambda = 1.05$ does not overwrite the parent state within the protocol window. The same weak field applied while $\lambda$ is ramped from $0.3$ to $1.05$ writes the new pattern cleanly, and the restored barriers then keep it. Ramping $\lambda$ without the field erases the parent pattern but writes no target (Fig.~\ref{fig:rewriting}d). At fixed high coupling the written overlap creeps upward only through slow activated escapes. Under a coupling ramp the rotors align while the barriers are low and are trapped once the coupling is restored.

Point-dipole arrays show the same activated behaviour on the target carrier. Weak patterned fields ($0.25$ and $0.75\,k_BT$) rewrite stored dipolar patterns by activation, and the logarithmic slopes against $\lambda_d$ approach the field-free retention slope as the field decreases (Fig.~\ref{fig:transfer}a). Rescaling the rotational mobility equally during writing and retention shifts both times together and leaves their ratio unchanged. Increasing it only during writing multiplies the ratio by the viscosity contrast. An example is writing orientational order into membrane nanorods while the lipid bilayer is fluid and retaining it after the bilayer has cooled into the gel phase. Each substrate gives the interventions of Sec.~\ref{sec:rewriting} a physical form. A genetic toggle or a chemical reactor is swept through its pitchfork or cusp. Collective barriers in colloidal or dipolar arrays are lowered during the write. Elastic linkages in geared metamaterials are softened, and the phase of a limit-cycle oscillator is shifted by a transient frequency pulse.

\section{Discussion}
\label{sec:discussion}

Comparing two routes between descriptions exposes a distinction that a single route can conceal. Commutators, connection curvature and loop holonomy are the algebraic, infinitesimal and finite forms of this comparison. On an interaction network, when the loop holonomy around a cycle, or across coupled cycles at a common base node, leaves no state invariant, the bond preferences cannot all be satisfied, and the network is frustrated into compromise configurations. Along a temporal protocol, the commutator of the write generators gives an order-dependent displacement of the state. This displacement records the sequence of operations whenever protocols that share the same final write still end in distinct basins. In coarse-graining, the commutator between the dynamical generator and the projection onto the measured observable identifies the excitation and the feedback return through which the eliminated coordinates generate a Mori--Zwanzig memory kernel. These coordinates store distinctions that a global order parameter does not show. Whether a path-dependent distinction survives as a memory, and what kinetic resistance opposes its replacement, is decided by the local restoring curvature $\kappa$ and by the barrier heights of the landscape along the carrying coordinate. Pairing the origin of the recorded distinction with the dynamical law of its loss arranges physical memories into Table~\ref{tab:classes}. It also shows why memory is a property of the complete realization $I_{\mathrm{real}}$ and not of a material substance or of a differential equation taken alone.

Several of the structural rules rest on classical results. Signed-graph balance and frustration functions decide whether bond preferences can be satisfied, and Thomas's feedback rules constrain the attractors of regulatory networks. Monotone systems and the Middleton no-passing theorem give conditions for return-point memory. Equivariant bifurcation theory identifies symmetric writing instabilities, and the Onsager--Kubo relations connect spontaneous relaxation to the forced response. To these, the present analysis adds the continuous loop-flux formula of Eq.~\eqref{eq:loop-flux}, which gives the geometric compatibility condition for reflection couplings that depend on the bond direction. It explains why the parity of odd and even cycles is inverted relative to antiferromagnets. Carrying that reflection structure from capillary colloidal rotors to point-dipole arrays shows that an orientational memory hidden from the net magnetization survives across different interaction ranges and loop fluxes. In the three-mode coupled system of Sec.~\ref{sec:selective-erasure}, a weak bias separates the two folds at which the mixed-sign patterns lose stability. A coupling pulse between the two thresholds erases one stored pattern and preserves the other, although the landscape before and after the pulse is the same.

Equation~\eqref{eq:ratio-bound} bounds the ratio of the mean retention time to the mean rewriting time under the specified static bias. Both the effective bias $E_{\mathrm{eff}}$ and the committor-weight factor $1+\delta$ enter the bound. In weakly tilted metastable wells with one dominant saddle, the leading barrier factor cancels from the ratio, and the Kramers--Bell dependence on the bias remains. A strong enough static bias can therefore give a large speed-up without a change of temperature or mobility; a large speed-up is consistent with the inequality at a large effective bias. When the available static bias is limited, modulation of the landscape and separate transition channels are the alternatives.

The two ways of writing differ in what the write has to overcome. Behind a formed barrier the write assists the rare fluctuation by which the state would in any case be lost, and it gains at most the Boltzmann factor of the energy that it supplies along that fluctuation. At an instability the coordinate has no restoring force for a time $T$, and the write is a detection of the bias by the coordinate itself. Its accuracy is $\Phi\bigl(h\sqrt{T/2D}\bigr)$, it is complete when the bias has done a work of a few $k_BT$ on the free coordinate, and the barrier enters only afterwards. In this way of writing, the state is chosen while the restoring force along the carrying coordinate vanishes, and it is retained by the barrier that forms afterwards. The restoring force within a state sets only the relaxation toward it; the height of the barrier and the noise set the escape. A phase on a limit cycle and a fading memory are retained without a barrier, by a flat direction or by a slow relaxation (Sec.~\ref{sec:loss-laws}). A memory that is both durable and rewritable needs a high barrier for retention and a free coordinate for writing, and meets the two requirements at different times, by a sweep, by heating or by a change of the coupling, or along different coordinates, as a photochromic switch does through its excited state. Landauer's principle concerns a different observable, the heat of logically irreversible erasure~\cite{Landauer1961}. The comparison made here concerns transition times and the bias that changes them.

The same holonomy describes the record that plastic deformation leaves in a crystal. In a perfect crystal a circuit of lattice steps closes; around a dislocation it fails to close. The closure failure, the Burgers vector, is the translational holonomy of the circuit, as the Frank vector of a disclination is the rotational one~\cite{Frank1951,KlemanFriedel2008}. In the continuum description the density of dislocations is the torsion of the lattice connection~\cite{Bilby1955}. A translation leaves no position unchanged, so the circuit cannot be satisfied, and the crystal accommodates the mismatch by elastic strain around the dislocation. The dislocations that remain after a plastic deformation therefore record it; those required by a gradient of the deformation record it through their net Burgers vector~\cite{Nye1953,Ashby1970}. The topology decides what such a record contains. How long the record lasts is decided, as for any other state, by the barriers that oppose the motion and annihilation of the defects (SI Sec.~\ref{si-sec:si-sources}).

The phase acquired by a quantum state carried around a loop of parameters~\cite{Berry1984} and the displacement of a microorganism whose changes of shape do not commute~\cite{ShapereWilczek1989} are holonomies of the kind considered here. A symmetry-breaking transition crossed at a finite rate is the write at an instability of Sec.~\ref{sec:write}, taking place in an extended system. Regions of a size set by the rate choose independently, and the topological defects left between them record the rate. The mechanism was proposed for cosmic strings~\cite{Kibble1976} and for superfluids~\cite{Zurek1985}, and observed in nematic liquid crystals~\cite{Chuang1991Cosmology}. The present description adds to the theory of geometric phases the question of how long a record lasts, which has an answer when the eliminated degrees of freedom act as a bath. The laws of loss require a closure in which the eliminated degrees of freedom act as a bath. In a closed system nothing is lost. A record dispersed by phase mixing moves into fine-grained coordinates, and a probe that couples them back to the observable, an echo, recovers it~\cite{Hahn1950}, as Sec.~\ref{sec:reading} requires of any measurement of a hidden record.

The required law of loss depends on the intended use of the record. A retention time $\tau$ requires barriers with $\Delta V/D \gtrsim \ln(\tau/\tau_0)$, a fading memory requires a controlled relaxation, and a flat direction keeps an analogue value only while its spread stays small compared with the separation of the written values. A memory that must be both durable and rewritable should be written where the landscape is unstable, or while its barrier is lowered or its mobility raised, and then retained behind the full barrier. The number of states and the retention time are set separately: the number by frustration and disorder, the retention time by the ratio of barrier to noise. A test that separates order from the final input compares three writes in orderings that share the final write, with a drive below saturation and a measurement after release. A record of the turning points of a drive is guaranteed when the elements can be relabeled so that the drive favours the same state of each element and every coupling favours equal states (SI Sec.~\ref{si-sec:si-design}).

Within a realization $I_{\mathrm{real}}$, engineering the loop holonomies and the disorder changes the compromise basins available to a write. Shaping the symmetries of a bifurcation, or the splitting of its folds, selects target states or erases them selectively. Modulating barriers, mobilities or excited-state channels separates long retention from rapid rewriting. To apply these interventions to a new substrate, one specifies its interactions, fluctuations and observables, and then calculates the writing protocol and the retained response.

\subsection*{Comparison with published measurements}

\begin{figure}[!tbp]
\centering
\includegraphics[width=0.62\linewidth]{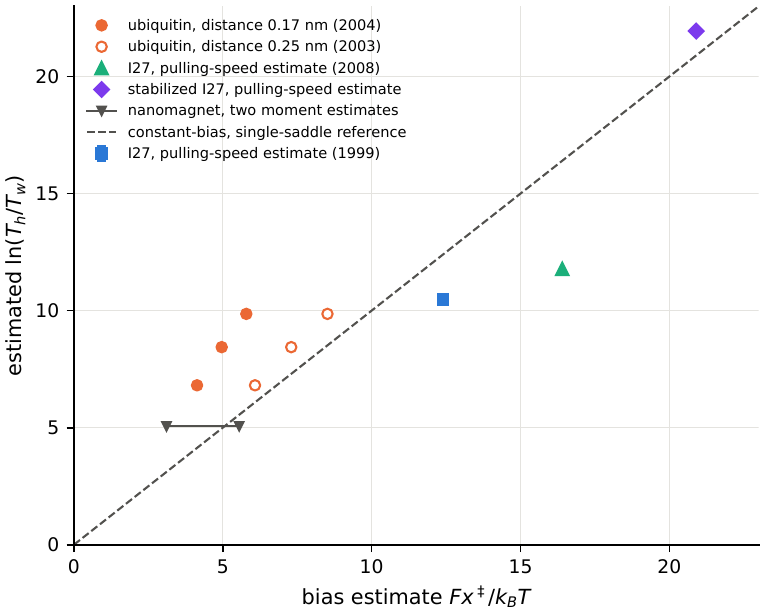}
\caption{\textbf{Published timescales compared with a constant-bias reference.} The logarithm of the estimated lifetime ratio is plotted against the bias estimate $F x^\ddagger$. The dashed line is the weak-noise single-saddle reference $\ln(\overline T_h/\overline T_w)=E_{\mathrm{eff}}/k_BT$ with $\delta=0$. It is not an exclusion line for measurements whose protocols or transition coordinates are not matched. Protein lifetimes without force come from chemical denaturation~\cite{Schlierf2004,CarrionVazquez1999,Borgia2008}. The ubiquitin points use fixed-force lifetimes at an assumed temperature of 298~K and two published distances~\cite{Schlierf2004,CarrionVazquez2003}; the forced lifetimes of I27 are estimated as the contour increment divided by the pulling speed. The nanomagnet points use rates from one fit and two estimates of the moment~\cite{Rippard2011}. All plotted quantities are calculated from the reported values.}
\label{fig:measurements}
\end{figure}

\paragraph{Retention against rewriting under a static bias} An experimental test of the exact mean-passage inequality of Eq.~\eqref{eq:ratio-bound} requires mean escape times measured with and without a static bias, at matched temperature and preparation, together with an independent determination of the effective energy bias across the unwritten transition region. Single-molecule force-spectroscopy and nanomagnet switching measurements use different loading histories and transition coordinates (Fig.~\ref{fig:measurements}; SI Sec.~\ref{si-sec:si-measurements}). In the mechanical unfolding of the titin I27 domain, chemical denaturation gives an independent lifetime at zero force. The forced lifetime can be estimated as the contour-length increment divided by the pulling speed. With this estimate the lifetime ratio lies about $2\,k_BT$ below the simple single-saddle reference line $\ln(\overline T_h/\overline T_w) = F x^\ddagger/k_BT$~\cite{CarrionVazquez1999}. Mutational stabilization of I27 reduces the zero-force unfolding rate by $2.6\times 10^4$, roughly $10\,k_BT$ of barrier height. It leaves the transition distance nearly unchanged ($0.33$ to $0.29$~nm) and raises the unfolding force at fixed pulling speed from $200$ to $290$~pN~\cite{Borgia2008}. The stiffer barrier slows spontaneous loss and forced unfolding together, and the stabilized protein lies about $1\,k_BT$ above the constant-bias reference line. The two variants share the estimated forced lifetime by construction, since both are the contour increment divided by the same pulling speed, so their comparison rests on the zero-force lifetimes and the forces; on this estimate the wild-type domain lies about $4.6\,k_BT$ below the line. Pulling at constant speed does not give a passage time at static force, so these points require a conversion for the loading history.

Force-clamp measurements on single ubiquitin molecules give lifetimes at fixed force directly~\cite{Schlierf2004}. With the transition distance $x^\ddagger = 0.17$~nm fitted over $100$--$140$~pN, the logarithm of the ratio of the zero-force chemical lifetime to the force-clamp lifetime lies $2.7$--$4.1\,k_BT$ above $F x^\ddagger/k_BT$. As Ref.~\cite{Schlierf2004} notes, however, the high-force rate extrapolated to zero force differs from the chemical unfolding rate by a factor of $40$--$100$. Either the transition pathway differs between chemical and mechanical unfolding, or the transition distance is larger ($\gtrsim 0.29$~nm) at low force, where the unwritten Dirichlet flux is concentrated. With the distance of $0.25$~nm from earlier constant-speed experiments~\cite{CarrionVazquez2003}, the same lifetimes lie $0.7$--$1.3\,k_BT$ above the line. Force-clamp measurements below $100$~pN under matched buffer conditions would distinguish between the two.

In telegraph-noise measurements on a nanoscale magnetic tunnel junction, a single Arrhenius fit gives an effective moment of $2.3\times 10^{-17}$~A\,m$^2$, while the moment of the physical volume is smaller by a factor of $1.8$~\cite{Rippard2011}. The escape rates and the coupling to the applied field must therefore be calibrated independently.

\paragraph{Writing through a separate channel} Diarylethene photoswitches write through an excited state. Their thermal half-lives in the ground state reach $470{,}000$~years, while optical switching completes in under $10$~ps~\cite{Irie2010}. The logarithm of the ratio of the two times is $56$. It is driven by the energy of a $365$~nm photon, $132\,k_BT$ at room temperature, and the transition passes between distinct electronic surfaces.

\paragraph{Writing at an instability} In a class-A semiconductor laser swept through threshold, the delayed onset scales with the modulation frequency with an exponent of $-0.50 \pm 0.01$~\cite{ElAmili2012}, which is the time scale $r^{-1/2}$ of Sec.~\ref{sec:write}. An optical parametric oscillator biased below one photon per pulse chooses its phase with a probability that is an error function of the bias~\cite{RoquesCarmes2023}, the form of Eq.~\eqref{eq:write-law}.

The selection itself was measured in an analogue electronic circuit swept through a pitchfork~\cite{KondepudiMossMcClintock1986}, and the same group gave Eq.~\eqref{eq:sweep-law} with its constant~\cite{KondepudiMossMcClintock1986Letter}. The law requires a small noise ratio $\Gamma = D|a_3|/r$, so that the state does not cross back over the ridge while the barrier grows. These papers report one operating point with $\Gamma < 1$, at $\Gamma = 0.25$. There the law gives a ratio of 4.4 between the two outcomes, against about 4.5 measured. The sweep rate was varied alone only at larger noise, by a factor of 30 from $\Gamma = 2$ to 62, and the argument of $\Phi$ then grows as $r^{-0.14}$ rather than $r^{-1/4}$. The Fokker--Planck equation of the cubic drift, solved at the published parameters, gives $r^{-0.12}$, because a slower sweep leaves more time to cross back (SI Sec.~\ref{si-sec:si-measurements}). At the published bias the Fokker--Planck solution lies 24--52\% below the measured ratios, and the stated offset of the circuit does not close the difference at a further operating point, $\Gamma = 2.6$. The circuit therefore agrees with the constant at one point and shows the condition on the noise. The exponent $1/4$ awaits a sweep at a small noise ratio, which is within reach of the same circuit. A Hamiltonian condensate crossed at a finite rate has a different exponent~\cite{Anquez2016}, so the law belongs to overdamped dynamics.

\paragraph{Laws of loss} A phase is lost by diffusion in a synthetic genetic oscillator: its squared error grows linearly with the number of periods, with a drift of 14\% per period~\cite{PotvinTrottier2016}. For the loss of a written pattern in a physical gel the predicted exponent is $1/3$; it has so far been measured by dynamic light scattering~\cite{Zhao2025GelHierarchy}, and its measurement on a written pattern is open to experiment.

\paragraph{Order of writes and structural rules} Mechanical metamaterials separate order from the last write in experiment. Of the six orders of three driving magnitudes, one drives a designed counter to a state that the other five do not reach, and the five include an order with the same last magnitude~\cite{Kwakernaak2023}. For the return to a turning point, serially coupled hysterons with antiferromagnetic-like interactions show pathways that break the return~\cite{Liu2024}, as the rule on loops through the drive (R8 in Table~\ref{tab:rules}) allows. The rule allows a failure but does not force one. In simulated amorphous solids, whose couplings have both signs, 2 of 84 transitions violate the return~\cite{Mungan2019}, and magnetic films whose macroscopic loops return perfectly never reach complete microscopic memory~\cite{Pierce2005}. An independent measurement of the signs of the couplings would test the rule directly. The flux of a loop of capillary or dipolar couplings is a further structural prediction open to experiment.

\subsection*{Scope and experimental tests}

The analytical laws of loss and the passage bounds derived here apply to overdamped Markovian dynamics with specified noise and boundary closures. The limit-cycle and transcriptional examples rest on centre-manifold and phase reductions of driven kinetics. In the colloidal and dipolar networks, the numerical tests use $N = 576$ rotors on five disordered graph geometries, with patterned fields aligned with target equilibria. Above $\lambda \approx 1$, partial trapping in nearby metastable basins reduces the apparent slope of the overlap decay. The return-point memory simulations use zero-temperature quasistatic driving on periodic square lattices and chains.

A measurement made after the field is removed, repeated at several temperatures, separates exponential relaxation, algebraic diffusion and Arrhenius activation. In interfacial colloidal monolayers, the target is to show three things: that a weak patterned field overwrites a stored orientational pattern with the same coupling activation as spontaneous escape, that a strong field removes the dependence on the barrier, and that ramping the coupling during a weak field pulse records patterns that cannot be written at fixed high coupling. In arrays of lithographic nanomagnets or of colloidal dipoles with fixed centres, the target is to show that bond-relative orientational patterns remain stored while the net magnetization vanishes. Near a chemical cusp, applying the compensated two-feed path of Eq.~\eqref{eq:schlogl-write-feed} tests selection at a pitchfork under a weak bias. Microfluidic or slime-mould networks with equal and unequal branch lengths test symmetric against biased threshold writing, together with the local stability of the unfavoured branch. In coupled mechanical or magnetic hysteron arrays, inverting at once the drive coupling and all interaction signs on a chosen element is a gauge relabeling that preserves exact return-point memory. Inverting either one alone frustrates the loops coupled to the drive and permits failures of the return; the target is to observe both cases. Applying the six permutations of three sub-saturating writes, grouped by a matched final input, tests order memory without frustration. Cooling nanorod-loaded lipid bilayers through their fluid--gel transition under a polarizing field~\cite{NanorodCompanion} tests writing on a modulated landscape, and tracking the decay of optical turbidity in dynamically crosslinked hydrogels~\cite{HyperionCompanion,Zhao2025GelHierarchy,Yu2020HydrogelMemory} tests stretched-exponential relaxation of a kernel on a soft-matter carrier.

A material that adjusts one of its parameters by comparing two realizations of its own response writes to that parameter. The two realizations can be two copies of the material compared at the same time, or the same material at two times. In the second case the earlier response must be retained until the comparison is made, and this is itself a memory of the kind described here. The parameters of the rule by which the material adjusts form a carrier one level higher. The description given here applies at each level. A tuned parameter without a restoring structure diffuses. A fixed barrier trades the retention of a tuned value against the speed of re-tuning, in the manner of Sec.~\ref{sec:rewriting}. A barrier that is lowered while successive comparisons agree is the counterpart of a landscape changed between writing and retention. These results, and the conditions that limit the depth of such a hierarchy, are the subject of the companion paper~\cite{SelfImprovementCompanion}.

\section*{Methods}

The overdamped Langevin equations were integrated with the Euler--Maruyama algorithm at time step $\Delta t = 2 \times 10^{-3}$. Each write condition used 4{,}000 independent trajectories. A binary write applied a field $h = \pm 1$ for a duration $1$, starting from the initial state $x = 0$.

\paragraph{Write laws} An instant write started Eq.~\eqref{eq:pitchfork} at $\epsilon = 1$ and $D = 0.02$ from the displaced state $x = s$, with $s = (WD/\epsilon)^{1/2}$ for $W = 1/16$, 1/4, 1 and 4. Each case was integrated over 12 time units at time step $2\times10^{-3}$ with 40{,}000 trajectories. The information kept by the well was computed as $\ln 2 - H_{\mathrm b}(p_{\mathrm{write}})$ and compared with the mutual information of a binary input in Gaussian noise at amplitude $\sqrt{W}$. A swept write integrated $\dot x = rtx - x^3 + h + \sqrt{2D}\,\xi$ from $t = -6/\sqrt{r}$ to $6/\sqrt{r}$ at $D = 0.01$ and time step $5\times10^{-3}$. The bias $h$ was chosen so that Eq.~\eqref{eq:sweep-law} predicts the probability $\Phi(1)$, and each sweep rate used 20{,}000 trajectories.

\paragraph{Retention and rewriting times} A bit was stored as the sign of $x$ in $\mathrm{d}x = [-V'(x) + f]\,\mathrm{d}t + \sqrt{2}\,\mathrm{d}B_t$, in units with $D = k_BT = 1$. The stored value was identified from $\operatorname{sign} x$, and the threshold probability of correct identification was $p^* = 0.9$. For a curved minimum, $V = \kappa x^2/2$, the state started from the stationary distribution under the force $\mp F$. For a flat direction, $V = 0$, it started at $\pm s_0$. For a barrier, $V = \Delta V (x^2 - 1)^2$, it started at $x = \pm 1$, and the tilt was $f = E_w$, equal to the first-order reduction of the barrier. The curved minima and the flat directions used 20{,}000 trajectories, and the barriers used 5{,}000. The Euler--Maruyama steps satisfied $\Delta t\,V''_{\min} \le 0.1$. The swept write used $D = 0.01$ and the sweep rate $r = 0.05$, starting from $\epsilon = -1$. Its bias was set by Eq.~\eqref{eq:sweep-law} for an accuracy of 0.99. The final values $\epsilon_f = 0.2$--1 give barriers $\epsilon_f^2/4D = 1$--25. The retention time at $\epsilon_f$ is the Kramers time over which the identification probability falls from the achieved accuracy to $p^*$. The rewrite on a fixed landscape applied the same bias to a state in the opposite well. It was simulated with 5{,}000 trajectories when it finished within about $10^3$ time units. Otherwise its time was estimated from Kramers' formula with the barrier lowered by $h\sqrt{\epsilon_f}$.

\paragraph{Colloidal rotors} The rotor network, the graphs and the integrator were those of Ref.~\cite{BaulinColloidMemory}: $N = 576$ rotors, couplings $(J, g) = \lambda\,(4, 5)\,k_BT$, a time step of $0.0025/D_r$ and five graphs. For each graph and coupling, 16 parent states were equilibrated for $125/D_r$ from random orientations, and each parent was split into two replicas. An independent state of the same graph and coupling, equilibrated in the same way, served as the new pattern. In the retention protocol the state evolved without a field. The rewriting protocols applied a field of energy $-h\cos 2(\theta_i - \theta_i^{B})$ per rotor. The field strengths were $h = 0.75$, 1.5 and $3\,k_BT$ for $\lambda = 0.3$--1.05. A second run with independent noise used $h = 0.25$ and $0.4\,k_BT$ for $\lambda = 0.6$--1.05, and in this range the two retention runs were averaged. The overlap with the stored state, $\langle\cos 2(\theta - \theta_{\mathrm{parent}})\rangle$, was recorded every $0.5/D_r$ for up to $2{,}000$--$3{,}000/D_r$. The escape time is the first time at which the replica-averaged overlap fell to half of its mean value between 0.3 and $0.7/D_r$. The slopes of $\ln T$ against $\lambda$ were fitted for each graph and then averaged over the graphs. The code is \texttt{scripts/rotating\_colloids\_hold\_rewrite\_test.py} and \texttt{simulations/analyze\_colloid\_hold\_rewrite.py}.

\paragraph{Writing while the coupling changes} The same five graphs were used, and the states were retained at the coupling $\lambda = 1.05$. For each graph, 16 parent states were equilibrated for $125/D_r$ from random orientations, and each parent was split into two replicas. One independently equilibrated state served as the new pattern. Four protocols acted for durations $T_w = 25$, 100 and $400/D_r$. The first applied a field of strength $h$ along the new pattern at $\lambda = 1.05$. The second applied the same field while $\lambda$ was set to 0.3 and then raised linearly back to 1.05 in 40 equal steps. The third applied the field at $\lambda = 0.3$ and was followed by an abrupt return to 1.05. The fourth raised $\lambda$ in the same way without a field. The field strengths were $h = 0.25$, 0.4 and $0.75\,k_BT$. After each protocol the field was removed and the state was kept at $\lambda = 1.05$ for $200/D_r$. The overlaps with the new pattern and with the parent were averaged over the last quarter of this interval, and then over replicas and graphs. The code is \texttt{scripts/rotating\_colloids\_sweep\_write\_test.py} and \texttt{simulations/analyze\_colloid\_sweep\_write.py}.

\paragraph{Dipole arrays} Dipoles placed at the positions of the five graphs were simulated with the rotor integrator through the map of Eq.~\eqref{eq:dipolar-map}. In this map the bond angles of the rotor model are set to half the geometric ones, every pair within 3.5 spacings carries the weight $(r_0/r)^3$, and the couplings are $J = \lambda_d/2$ and $g = 3\lambda_d/2$. The map reproduces the directly evaluated dipolar energy of random configurations to a relative error below $10^{-14}$. In the rotor variable the dipole moment diffuses with $4D_r$, and times are given in units of $1/D_r$. The retention and rewriting protocols of the colloid test were run for $\lambda_d = 1$--6, with fields of 0.25 and $0.75\,k_BT$ and records of up to $2{,}000/D_r$. The escape times were defined as for the colloids. The order of the moments relative to their bonds was measured as the mean of $\cos 2(\psi_i + \psi_j - 2\varphi_{ij})$ over the bonds in the equilibrated states. The code is \texttt{scripts/rotating\_colloids\_dipolar\_transfer\_test.py} and \texttt{simulations/analyze\_dipolar\_transfer.py}.

\paragraph{Construction on other carriers} Each realization was written as a drift on its carrier (a torus of $m$-fold angles, a concentration orthant or $\mathbb{R}^n$), with a control parameter and, where relevant, further material parameters. The code is the memory module of FieldBridge~\cite{FieldBridge} (\texttt{analysis.py}, \texttt{construct.py}, \texttt{compose.py}, \texttt{library.py}), with the command-line interface \texttt{fieldbridge memory}. Stable states were found by relaxation from random starts, which were uniform and log-uniform in concentration. Write points were located by predictor--corrector continuation of every stable state between neighbouring values of the control. The continuation stopped where an eigenvalue of the Jacobian crossed zero, and this crossing was refined by bisection, or where the state ceased to exist. The steps were halved down to $2^{-18}$ of the interval, and states on the boundary of an orthant were not continued. At a write point, eigenvalues whose real part lay within the residual of the critical eigenvalue plus its change over $2 \times 10^{-3}$ of the scanned range were counted as critical together. A write point with more than one critical direction (more than a complex pair at a Hopf point) would be reported as a degenerate critical mode and would not be reduced. Every write point of Table~\ref{tab:write-points} has a single critical direction or pair. The critical right and left eigenvectors of the Jacobian at the write point define the coordinate $u$. For each $u$ on a grid of 41 points, the transverse coordinates were solved so that the drift had no transverse component, and the drift along $u$ was fitted by a cubic. The quadratic term counts as an obstruction when $|a_2|u_{\max}^2 > 0.1\,(|a_2|u_{\max}^2 + |a_3|u_{\max}^3)$. The bias counts when the component of $\partial_v\Omega$ along the critical direction exceeds 0.1 of the sum of this component and $|\partial_v a_1|\,u_{\max}$. Here $\partial_v a_1$ is taken along the moving state, $\mathrm{d}q/\mathrm{d}v = -\mathcal J^{+}\partial_v\Omega$. Cusps were found by following the fold, $\Omega = 0$ and $\det\mathcal J = 0$, in the second parameter until $a_2$ changed sign, and were refined with the condition $a_2 = 0$ added. The swept writes used 1{,}500 trajectories, a bias of fixed size along the critical direction and a sweep at $r = 0.02$ from $4r^{-1/2}$ before to $4r^{-1/2}$ after the critical value. The genetic toggle has Hill exponent 2, which gives $\alpha_c = \ToggleAlphaC$. Loops were simulated from 32 starts. The genes of the loops and rings had Hill exponent 4 and $\alpha = 10$, and the soft spins had $\epsilon = 1$ and $k = 0.4$. The rotors had unit bond strengths, and their positions were displaced from a regular polygon by Gaussian noise of 0.3 bond lengths. The adaptive tubes follow $\dot G_k = f(|Q_k|) - G_k$ with $f(Q) = Q^{n_{\mathrm H}}/(1 + Q^{n_{\mathrm H}})$ at a fixed total flux. The numbers of Table~\ref{tab:write-points} and of the single loops are generated from this module by \texttt{simulations/analyze\_constructor.py}.

\paragraph{Structural predictions and networks} Structural predictions were obtained from a model without integrating it. The signs of the interaction graph were taken from the off-diagonal entries of the Jacobian on sampled states, and reciprocity from the symmetry of the Jacobian. The symmetries of the drift are the permutations of the variables, combined with sign reversals or reflections of angles, that commute with the drift. The continuous symmetries are the directions along which the drift does not change. The role of the control parameter was determined from its action on the drift. For networks the transports of the edges were used directly. Random networks of four to eight units consisted of a random spanning tree and between one and $n - 1$ further edges. Rotor networks had positions uniform in a square, and their bonds were reflections, alignments or anti-alignments with probabilities 0.6, 0.25 and 0.15. Spin networks had antiferromagnetic bonds with probability 0.4. Gene networks had one or two regulators per gene, with random signs or, in one network out of three, with signs chosen so that every undirected cycle was positive. Each network was relaxed from 32 random states. For rotors the prediction was compared with the lowest energy found and with the number of zero modes of the lowest state. For spins it was compared with the unsatisfied bonds of the lowest state. For genes it was compared with the number of stable states and with the presence of starts that did not settle. The phase of the ring of three repressors is described in SI Sec.~\ref{si-sec:si-phase}. The code is the memory module of FieldBridge~\cite{FieldBridge} (\texttt{predict.py}, \texttt{networks.py}, \texttt{phase.py}), run by \texttt{simulations/analyze\_predictions.py}.

\paragraph{Order of writes} The ring of eight coordinates and the three compartments are specified in SI Secs.~\ref{si-sec:si-order} and~\ref{si-sec:si-compartments}. The ring was integrated by an adaptive method with relative tolerance $10^{-10}$. After the field-free relaxation, a state was accepted as retained when a Newton iteration converged to a stationary point of the energy with a positive definite Hessian, at a distance below $10^{-4}$. The compartment model was evaluated with matrix exponentials. The numbers are generated by \texttt{simulations/analyze\_order.py}.

\paragraph{Return to a turning point} The elements $s_i = \pm 1$ sit on a periodic square lattice of $\RpmL \times \RpmL$ sites. They interact through nearest-neighbour couplings $J_{ij} = \pm 1$ and feel random fields $h_i$, drawn from a Gaussian of width $\sigma_h = \RpmR$, and the drive $\eta_iH$. An element is stable when $s_i(\sum_j J_{ij}s_j + h_i + \eta_iH) > 0$. The drive was moved from one instability to the next without thermal noise. At each instability the avalanche was relaxed at fixed $H$ by flipping the least stable element, or, in a control, a randomly chosen unstable element, until every element was stable. The drive started from a large negative value, at which every element is in the state that the drive favours. It then rose to $H_1$, fell to $H_2 < H_1$ and rose to $H_1$ again. The subloop returned if the configuration at the second arrival at $H_1$ equalled the configuration at the first arrival. Each condition used \RpmSamples\ samples with six subloops each. The value $H_1$ was uniform between $\pm 0.8\sigma_h$, and the difference $H_1 - H_2$ was uniform between $0.3\sigma_h$ and $1.5\sigma_h$. Chains of \RpmChainL\ elements were simulated in the same way. The code is \texttt{simulations/sim\_return\_point\_memory.py} (SI Sec.~\ref{si-sec:si-return}).

\paragraph{Lifetime of a stable state} The lifetime of Sec.~\ref{sec:loss-laws} was computed from the equations of each model record at its declared parameters, with stable states and fixed points found from fixed random starts (SI Sec.~\ref{si-sec:si-lifetime}). A state is predicted kept when the fraction lost during the hold is below 0.1, the threshold of the hold check of the records. For the test, 3{,}000 trajectories were integrated from each stable state by the Euler--Maruyama method with the declared noise over the hold $T=56.6$ of the records, as long as the write, and a trajectory that ends nearer to another stable state or diverges counts as lost.

\paragraph{Fields} The field simulations are described in SI Sec.~\ref{si-sec:si-fields}.

\paragraph{Selective erasure} The model of Eq.~\eqref{eq:selective-model} was analysed at $\varrho=0.16$, with the bias $h_1=0.001580408$. The coupling interval follows from the fold equations and the Jacobian spectra. The folds lie at $\chi_-=0.02795253$ and $\chi_+=0.02929186$, and the pulse amplitude is their midpoint, $\chi=0.02862219$. At each threshold the equilibrium residual vanishes, a single eigenvalue of the Jacobian crosses zero while the other two remain negative, and both the quadratic coefficient of the normal form and the transversality in the parameter are non-zero. The pulse duration is twice the local estimate of the passage time. The finite-noise calculation uses $D=1.0708793\times10^{-7}$ and 256 reference trajectories from each prescribed state. The settings were saved before the reference trajectories were evaluated, the noise was paired across protocols, and the integration step was refined. Weaker, stronger and shorter pulses serve as controls. Independent DOP853 and Radau calculations check the deterministic endpoint map (SI Sec.~\ref{si-sec:si-selective}). The stochastic results come from the archived trajectories, and the fold calculation is reproduced independently.

\paragraph{Published models} The survey used FieldBridge records~\cite{FieldBridge} of 134 supplied specifications, 103 compiled from arXiv sources and 31 of hardware. Two conditions, each an exact property of a specification, select the models that are counted. A compiled specification with three or more parameters, all equal to one, carries no parameter values of its source, and the position of its instabilities is then a statement about placeholder values; 13 specifications are of this kind. Specifications whose drifts have the same form up to the names of variables and parameters are copies of one system, and one copy is kept, the one whose control covers the widest range; 11 specifications are copies, four of them of Lorenz's model. The survey therefore counts 110 models. A cusp along a second parameter was counted only when it lies at a value that the parameter can take: of 89 cusps, five lie at zero or at the opposite sign of a rate, and they were not counted. The survey followed the stable states along each declared control and reduced the critical directions at their instabilities. Connected sets of equilibria were merged, and directions that remained neutral on both sides of an instability were not counted as crossings. The completion step retained degenerate critical spaces, followed folds to cusps, kept higher saturating terms or followed distant branches. The original swept tests used 2,000 trajectories per bias at $D_s|a_3|/r=0.05$ where a cubic was resolved. Each sweep was followed by an interval as long as the sweep, with the control kept at its final value. Steps that took more than four hours remained unresolved. The high incidence of write instabilities among the 31 hardware models reflects this engineering selection and is not an unbiased prevalence across theoretical models. The search for write points also stopped with a numerical error in eight models, and these cases do not show that the models lack instabilities. The corrected hardware records recalculate the outcome map separately and follow the slow directions over the writing timescale. They compare the reduced probability law with the full stochastic evolution and distinguish an unfinished choice from a reached alternative. They then measure the changes of state during the interval in which the control is kept at its final value. Three settled outcomes are recorded separately, and a sweep whose modes have not settled is recorded as not settled: a selection that agrees with the reduced law of the linear stage, a choice made in the nonlinear stage and retained, and continuing equilibrium exchange (SI Sec.~\ref{si-sec:si-published}).

\paragraph{Numbers and code} The values of existing simulations are read from stored results through \texttt{numbers.tex}. The original aggregate point survey and the corrected hardware writing records of 31 models are identified separately in SI Sec.~\ref{si-sec:si-published}. The corrected hardware counts are read independently from the archived JSONL records of FieldBridge at commit 94ab84d. The accompanying script \texttt{check\_mathematics.py} checks the matrix identities, the passage identities of finite reversible networks and the folds of the selective-erasure model. Its calculation receipt lists the source hashes and the numerical residuals. The additional checks in \texttt{check\_corrections.py} cover the passage-time examples, the transport orientations and the compensated chemical feed.
\section*{Data and code availability}
FieldBridge~\cite{FieldBridge}, which contains the memory module used for the constructions on other carriers, the structural predictions, the networks and the records of the published models, is publicly available under the PolyForm Noncommercial License 1.0.0. The simulation scripts for the write laws, the retention and rewriting times, the colloidal rotors, the dipole arrays, the order of writes, the return to a turning point and the lifetime of stable states, with their random seeds and stored results, and the specifications and records of the 134 published models are available from the corresponding author on request.

\section*{CRediT authorship contribution statement}
\textbf{Vladimir A. Baulin:} Conceptualization, Methodology, Software, Formal analysis, Investigation, Writing -- original draft, Writing -- review \& editing.

\section*{Declaration of competing interest}
The author has no known competing financial interests or personal relationships that could have appeared to influence the work reported in this paper.

\section*{Declaration of generative AI and AI-assisted technologies in the manuscript preparation process}
During the preparation of this work the author used Claude (Anthropic) to assist with programming, numerical analysis and the preparation of the text. After using this tool, the author reviewed and edited the content as needed and takes full responsibility for the content of the publication.

\section*{Acknowledgments}
The constructions on other carriers and the records of the published models were made with FieldBridge~\cite{FieldBridge}.

\section*{Funding}
This research did not receive any specific grant from funding agencies in the public, commercial, or not-for-profit sectors.

\clearpage
\includepdf[pages=-]{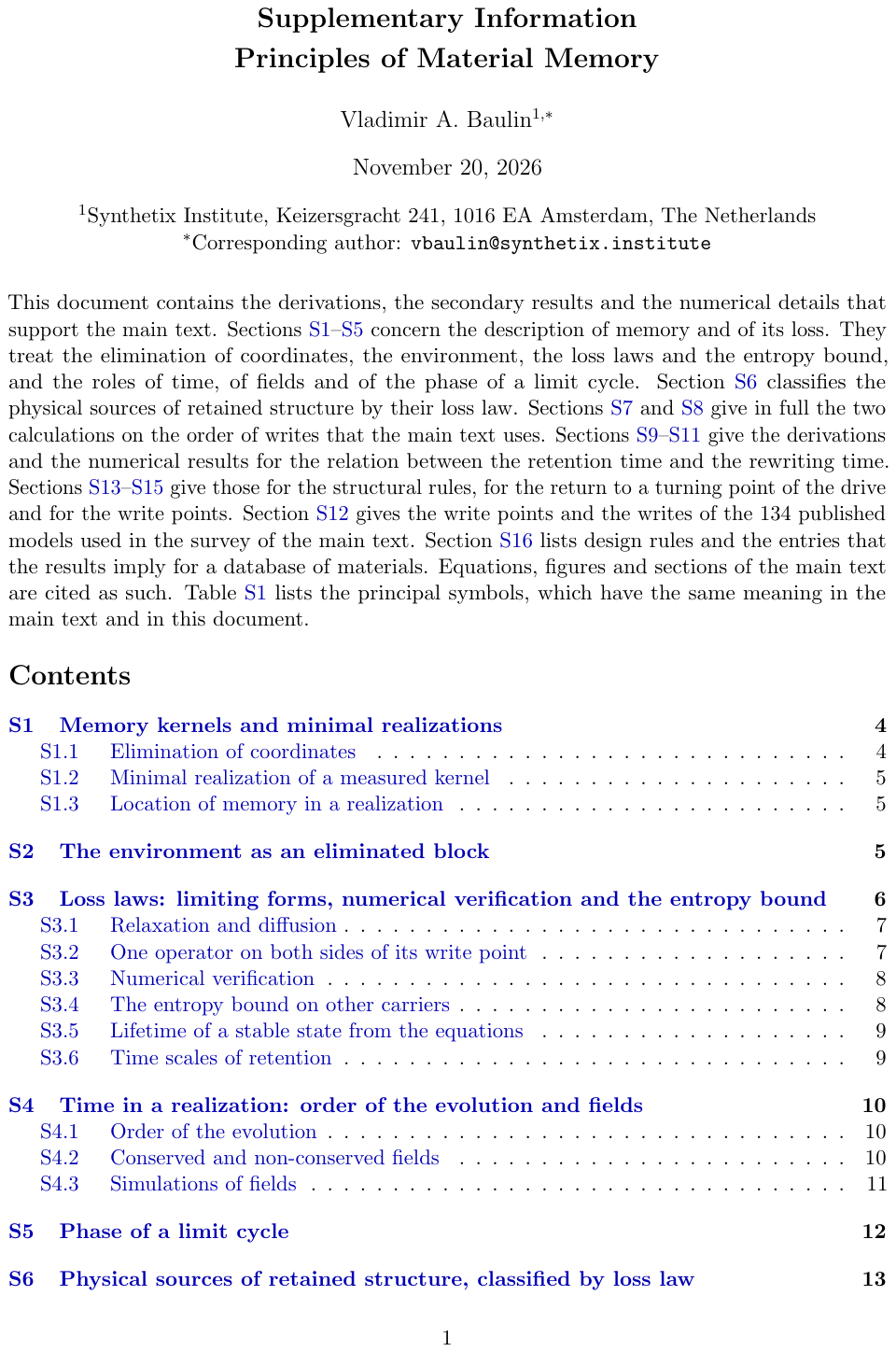}

\begin{thebibliography}{99}
\bibitem{Keim2019Memory} N. C. Keim, J. D. Paulsen, Z. Zeravcic, S. Sastry and S. R. Nagel, Memory formation in matter, \emph{Rev. Mod. Phys.} \textbf{91}, 035002 (2019).
\bibitem{PaulsenKeim2025} J. D. Paulsen and N. C. Keim, Mechanical memories in solids, from disorder to design, \emph{Annu. Rev. Condens. Matter Phys.} \textbf{16}, 61--81 (2025).
\bibitem{Barkhausen1919} H. Barkhausen, Zwei mit Hilfe der neuen Verst\"arker entdeckte Erscheinungen, \emph{Phys. Z.} \textbf{20}, 401--403 (1919).
\bibitem{Preisach1935} F. Preisach, \"Uber die magnetische Nachwirkung, \emph{Z. Phys.} \textbf{94}, 277--302 (1935).
\bibitem{Sethna1993Hysteresis} J. P. Sethna, K. Dahmen, S. Kartha, J. A. Krumhansl, B. W. Roberts and J. D. Shore, Hysteresis and hierarchies: dynamics of disorder-driven first-order phase transformations, \emph{Phys. Rev. Lett.} \textbf{70}, 3347--3350 (1993).
\bibitem{Kovacs1979} A. J. Kovacs, J. J. Aklonis, J. M. Hutchinson and A. R. Ramos, Isobaric volume and enthalpy recovery of glasses. II. A transparent multiparameter theory, \emph{J. Polym. Sci. Polym. Phys. Ed.} \textbf{17}, 1097--1162 (1979).
\bibitem{Bouchaud1992Aging} J.-P. Bouchaud, Weak ergodicity breaking and aging in disordered systems, \emph{J. Phys. I} \textbf{2}, 1705--1713 (1992).
\bibitem{Cugliandolo1993Aging} L. F. Cugliandolo and J. Kurchan, Analytical solution of the off-equilibrium dynamics of a long-range spin-glass model, \emph{Phys. Rev. Lett.} \textbf{71}, 173--176 (1993).
\bibitem{Corte2008Reversible} L. Cort\'e, P. M. Chaikin, J. P. Gollub and D. J. Pine, Random organization in periodically driven systems, \emph{Nat. Phys.} \textbf{4}, 420--424 (2008).
\bibitem{Keim2011Memory} N. C. Keim and S. R. Nagel, Generic transient memory formation in disordered systems with noise, \emph{Phys. Rev. Lett.} \textbf{107}, 010603 (2011).
\bibitem{Paulsen2014Memory} J. D. Paulsen, N. C. Keim and S. R. Nagel, Multiple transient memories in experiments on sheared non-Brownian suspensions, \emph{Phys. Rev. Lett.} \textbf{113}, 068301 (2014).
\bibitem{Matan2002Crumpling} K. Matan, R. B. Williams, T. A. Witten and S. R. Nagel, Crumpling a thin sheet, \emph{Phys. Rev. Lett.} \textbf{88}, 076101 (2002).
\bibitem{Yu2020HydrogelMemory} C. Yu \emph{et al.}, Hydrogels as dynamic memory with forgetting ability, \emph{Proc. Natl. Acad. Sci. U.S.A.} \textbf{117}, 18962--18968 (2020).
\bibitem{Zhao2025GelHierarchy} Y. H. Zhao \emph{et al.}, Universal law of hierarchical dynamics in gels arising from confluence of local physically dynamic bonds, \emph{Nat. Commun.} \textbf{16}, 3247 (2025).
\bibitem{Silverberg2014Origami} J. L. Silverberg \emph{et al.}, Using origami design principles to fold reprogrammable mechanical metamaterials, \emph{Science} \textbf{345}, 647--650 (2014).
\bibitem{Florijn2014Metamaterials} B. Florijn, C. Coulais and M. van Hecke, Programmable mechanical metamaterials, \emph{Phys. Rev. Lett.} \textbf{113}, 175503 (2014).
\bibitem{Coulais2016Metamaterials} C. Coulais, E. Teomy, K. de Reus, Y. Shokef and M. van Hecke, Combinatorial design of textured mechanical metamaterials, \emph{Nature} \textbf{535}, 529--532 (2016).
\bibitem{Chen2021Reprogrammable} T. Chen, M. Pauly and P. M. Reis, A reprogrammable mechanical metamaterial with stable memory, \emph{Nature} \textbf{589}, 386--390 (2021).
\bibitem{Chua1971Memristor} L. O. Chua, Memristor---the missing circuit element, \emph{IEEE Trans. Circuit Theory} \textbf{18}, 507--519 (1971).
\bibitem{Strukov2008Memristor} D. B. Strukov, G. S. Snider, D. R. Stewart and R. S. Williams, The missing memristor found, \emph{Nature} \textbf{453}, 80--83 (2008).
\bibitem{Yang2013Memristor} J. J. Yang, D. B. Strukov and D. R. Stewart, Memristive devices for computing, \emph{Nat. Nanotechnol.} \textbf{8}, 13--24 (2013).
\bibitem{Mayergoyz1991} I. D. Mayergoyz, \emph{Mathematical Models of Hysteresis} (Springer, New York, 1991).
\bibitem{Arnold1989} V. I. Arnold, \emph{Mathematical Methods of Classical Mechanics}, 2nd edn (Springer, New York, 1989).
\bibitem{KobayashiNomizu1996} S. Kobayashi and K. Nomizu, \emph{Foundations of Differential Geometry}, Vol. I (Wiley, New York, 1996; first published 1963).
\bibitem{HyperionCompanion} V. A. Baulin, A. Cook and S. Rahman, Constructing physical mechanisms with artificial intelligence, manuscript in preparation (2026).
\bibitem{StructuralComplexityCompanion} V. A. Baulin, Thermodynamic limit of structural complexity in matter, manuscript in preparation (2026).
\bibitem{SelfImprovementCompanion} V. A. Baulin, Self-improvement in materials: comparison, retention and the depth of recursion, manuscript in preparation (2026).
\bibitem{TeitelJayaprakash1983} S. Teitel and C. Jayaprakash, Phase transitions in frustrated two-dimensional XY models, \emph{Phys. Rev. B} \textbf{27}, 598--601 (1983).
\bibitem{Toulouse1977} G. Toulouse, Theory of the frustration effect in spin glasses: I, \emph{Commun. Phys.} \textbf{2}, 115--119 (1977).
\bibitem{Mori1965} H. Mori, Transport, collective motion, and Brownian motion, \emph{Prog. Theor. Phys.} \textbf{33}, 423--455 (1965).
\bibitem{Zwanzig1961} R. Zwanzig, Memory effects in irreversible thermodynamics, \emph{Phys. Rev.} \textbf{124}, 983--992 (1961).
\bibitem{Zwanzig2001} R. Zwanzig, \emph{Nonequilibrium Statistical Mechanics} (Oxford University Press, 2001).
\bibitem{Zwanzig1973} R. Zwanzig, Nonlinear generalized Langevin equations, \emph{J. Stat. Phys.} \textbf{9}, 215--220 (1973).
\bibitem{Meyer2017NonStationary} H. Meyer, T. Voigtmann and T. Schilling, On the non-stationary generalized Langevin equation, \emph{J. Chem. Phys.} \textbf{147}, 214110 (2017).
\bibitem{HoKalman1966} B. L. Ho and R. E. Kalman, Effective construction of linear state-variable models from input/output functions, \emph{Regelungstechnik} \textbf{14}, 545--548 (1966).
\bibitem{BoydChua1985} S. Boyd and L. O. Chua, Fading memory and the problem of approximating nonlinear operators with Volterra series, \emph{IEEE Trans. Circuits Syst.} \textbf{32}, 1150--1171 (1985).
\bibitem{Gardner2000Toggle} T. S. Gardner, C. R. Cantor and J. J. Collins, Construction of a genetic toggle switch in \emph{Escherichia coli}, \emph{Nature} \textbf{403}, 339--342 (2000).
\bibitem{Kryder2008HAMR} M. H. Kryder, E. C. Gage, T. W. McDaniel, W. A. Challener, R. E. Rottmayer, G. Ju, Y.-T. Hsia and M. F. Erden, Heat assisted magnetic recording, \emph{Proc. IEEE} \textbf{96}, 1810--1835 (2008).
\bibitem{Moessner2006} R. Moessner and A. P. Ramirez, Geometrical frustration, \emph{Phys. Today} \textbf{59}, 24--29 (2006).
\bibitem{Wang2006SpinIce} R. F. Wang \emph{et al.}, Artificial `spin ice' in a geometrically frustrated lattice of nanoscale ferromagnetic islands, \emph{Nature} \textbf{439}, 303--306 (2006).
\bibitem{BaulinColloidMemory} V. A. Baulin, Geometry-encoded hidden orientational memory, arXiv:2610.19090 (2026).
\bibitem{SequenceCompanion} V. A. Baulin, Matter that remembers an order, and acts on it, manuscript in preparation (2026).
\bibitem{Croce2026} N. Croce, H. Salahshoor and D. Z. Rocklin, Multi-field return point memory, arXiv:2605.23781 (2026).
\bibitem{Elowitz2000Repressilator} M. B. Elowitz and S. Leibler, A synthetic oscillatory network of transcriptional regulators, \emph{Nature} \textbf{403}, 335--338 (2000).
\bibitem{Nye1953} J. F. Nye, Some geometrical relations in dislocated crystals, \emph{Acta Metall.} \textbf{1}, 153--162 (1953).
\bibitem{Ashby1970} M. F. Ashby, The deformation of plastically non-homogeneous materials, \emph{Philos. Mag.} \textbf{21}, 399--424 (1970).
\bibitem{Garanin2022RandomAnisotropy} D. A. Garanin and E. M. Chudnovsky, Random anisotropy magnet at finite temperature, \emph{J. Phys.: Condens. Matter} \textbf{34}, 285801 (2022).
\bibitem{FisherHuse1988} D. S. Fisher and D. A. Huse, Nonequilibrium dynamics of spin glasses, \emph{Phys. Rev. B} \textbf{38}, 373 (1988).
\bibitem{Jonason1998Memory} K. Jonason, E. Vincent, J. Hammann, J.-P. Bouchaud and P. Nordblad, Memory and chaos effects in spin glasses, \emph{Phys. Rev. Lett.} \textbf{81}, 3243 (1998).
\bibitem{Bouchaud2001Microscope} J.-P. Bouchaud, V. Dupuis, J. Hammann and E. Vincent, Separation of time and length scales in spin glasses: temperature as a microscope, \emph{Phys. Rev. B} \textbf{65}, 024439 (2001).
\bibitem{AmirOregImry2012} A. Amir, Y. Oreg and Y. Imry, On relaxations and aging of various glasses, \emph{Proc. Natl. Acad. Sci. U.S.A.} \textbf{109}, 1850--1855 (2012).
\bibitem{SethnaDahmenMyers2001} J. P. Sethna, K. A. Dahmen and C. R. Myers, Crackling noise, \emph{Nature} \textbf{410}, 242--250 (2001).
\bibitem{KatzgraberYoung2002} H. G. Katzgraber and A. P. Young, Numerical studies of the two- and three-dimensional gauge glass at low temperature, \emph{Phys. Rev. B} \textbf{66}, 224507 (2002).
\bibitem{KondepudiNelson1985} D. K. Kondepudi and G. W. Nelson, Weak neutral currents and the origin of biomolecular chirality, \emph{Nature} \textbf{314}, 438--441 (1985).
\bibitem{KondepudiMossMcClintock1986} D. K. Kondepudi, F. Moss and P. V. E. McClintock, Observation of symmetry breaking, state selection and sensitivity in a noisy electronic system, \emph{Physica D} \textbf{21}, 296--306 (1986).
\bibitem{KondepudiMossMcClintock1986Letter} D. K. Kondepudi, F. Moss and P. V. E. McClintock, Branch selection in the presence of coloured noise, \emph{Phys. Lett. A} \textbf{114}, 68--74 (1986).
\bibitem{Kramers1940} H. A. Kramers, Brownian motion in a field of force and the diffusion model of chemical reactions, \emph{Physica} \textbf{7}, 284--304 (1940).
\bibitem{Hanggi1990} P. H\"anggi, P. Talkner and M. Borkovec, Reaction-rate theory: fifty years after Kramers, \emph{Rev. Mod. Phys.} \textbf{62}, 251--341 (1990).
\bibitem{Gross1975} L. Gross, Logarithmic Sobolev inequalities, \emph{Am. J. Math.} \textbf{97}, 1061--1083 (1975).
\bibitem{BakryEmery1985} D. Bakry and M. \'Emery, Diffusions hypercontractives, in \emph{S\'eminaire de Probabilit\'es XIX}, Lecture Notes in Mathematics \textbf{1123}, pp.~177--206 (Springer, Berlin, 1985).
\bibitem{OttoVillani2000} F. Otto and C. Villani, Generalization of an inequality by Talagrand and links with the logarithmic Sobolev inequality, \emph{J. Funct. Anal.} \textbf{173}, 361--400 (2000).
\bibitem{MenzSchlichting2014} G. Menz and A. Schlichting, Poincar\'e and logarithmic Sobolev inequalities by decomposition of the energy landscape, \emph{Ann. Probab.} \textbf{42}, 1809--1884 (2014).
\bibitem{DiaconisSaloffCoste1996} P. Diaconis and L. Saloff-Coste, Logarithmic Sobolev inequalities for finite Markov chains, \emph{Ann. Appl. Probab.} \textbf{6}, 695--750 (1996).
\bibitem{ErbarMaas2012} M. Erbar and J. Maas, Ricci curvature of finite Markov chains via convexity of the entropy, \emph{Arch. Ration. Mech. Anal.} \textbf{206}, 997--1038 (2012).
\bibitem{KastoryanoTemme2013} M. J. Kastoryano and K. Temme, Quantum logarithmic Sobolev inequalities and rapid mixing, \emph{J. Math. Phys.} \textbf{54}, 052202 (2013).
\bibitem{CarlenMaas2017} E. A. Carlen and J. Maas, Gradient flow and entropy inequalities for quantum Markov semigroups with detailed balance, \emph{J. Funct. Anal.} \textbf{273}, 1810--1869 (2017).
\bibitem{CoverThomas2006} T. M. Cover and J. A. Thomas, \emph{Elements of Information Theory}, 2nd edn (Wiley, 2006).
\bibitem{HohenbergHalperin1977} P. C. Hohenberg and B. I. Halperin, Theory of dynamic critical phenomena, \emph{Rev. Mod. Phys.} \textbf{49}, 435--479 (1977).
\bibitem{Onsager1931} L. Onsager, Reciprocal relations in irreversible processes. II, \emph{Phys. Rev.} \textbf{38}, 2265--2279 (1931).
\bibitem{Kubo1966} R. Kubo, The fluctuation-dissipation theorem, \emph{Rep. Prog. Phys.} \textbf{29}, 255--284 (1966).
\bibitem{Bell1978} G. I. Bell, Models for the specific adhesion of cells to cells, \emph{Science} \textbf{200}, 618--627 (1978).
\bibitem{Bovier2004Metastability} A. Bovier, M. Eckhoff, V. Gayrard and M. Klein, Metastability in reversible diffusion processes I. Sharp asymptotics for capacities and exit times, \emph{J. Eur. Math. Soc.} \textbf{6}, 399--424 (2004).
\bibitem{Fusi2005Cascade} S. Fusi, P. J. Drew and L. F. Abbott, Cascade models of synaptically stored memories, \emph{Neuron} \textbf{45}, 599--611 (2005).
\bibitem{BennaFusi2016} M. K. Benna and S. Fusi, Computational principles of synaptic memory consolidation, \emph{Nat. Neurosci.} \textbf{19}, 1697--1706 (2016).
\bibitem{Irie2000Diarylethenes} M. Irie, Diarylethenes for memories and switches, \emph{Chem. Rev.} \textbf{100}, 1685--1716 (2000).
\bibitem{Harary1953} F. Harary, On the notion of balance of a signed graph, \emph{Michigan Math. J.} \textbf{2}, 143--146 (1953).
\bibitem{Nussinov2015Compass} Z. Nussinov and J. van den Brink, Compass models: Theory and physical motivations, \emph{Rev. Mod. Phys.} \textbf{87}, 1--59 (2015).
\bibitem{Loudet2005} J. C. Loudet, A. M. Alsayed, J. Zhang and A. G. Yodh, Capillary interactions between anisotropic colloidal particles, \emph{Phys. Rev. Lett.} \textbf{94}, 018301 (2005).
\bibitem{Stamou2000} D. Stamou, C. Duschl and D. Johannsmann, Long-range attraction between colloidal spheres at the air-water interface: The consequence of an undulated contact line, \emph{Phys. Rev. E} \textbf{62}, 5263--5272 (2000).
\bibitem{Bowden1997} N. Bowden, A. Terfort, J. Carbeck and G. M. Whitesides, Self-assembly of mesoscale objects into ordered two-dimensional arrays, \emph{Science} \textbf{276}, 233--235 (1997).
\bibitem{Pauling1935} L. Pauling, The structure and entropy of ice and of other crystals with some randomness of atomic arrangement, \emph{J. Am. Chem. Soc.} \textbf{57}, 2680--2684 (1935).
\bibitem{Thomas1981} R. Thomas, On the relation between the logical structure of systems and their ability to generate multiple steady states or sustained oscillations, in \emph{Numerical Methods in the Study of Critical Phenomena}, Springer Series in Synergetics \textbf{9}, 180--193 (Springer, Berlin, 1981).
\bibitem{Soule2003} C. Soul\'e, Graphic requirements for multistationarity, \emph{ComPlexUs} \textbf{1}, 123--133 (2003).
\bibitem{Snoussi1998} E. H. Snoussi, Necessary conditions for multistationarity and stable periodicity, \emph{J. Biol. Syst.} \textbf{6}, 3--9 (1998).
\bibitem{Golubitsky2002} M. Golubitsky and I. Stewart, \emph{The Symmetry Perspective} (Birkh\"auser, Basel, 2002).
\bibitem{Middleton1992} A. A. Middleton, Asymptotic uniqueness of the sliding state for charge-density waves, \emph{Phys. Rev. Lett.} \textbf{68}, 670--673 (1992).
\bibitem{vanHecke2021Hysterons} M. van Hecke, Profusion of transition pathways for interacting hysterons, \emph{Phys. Rev. E} \textbf{104}, 054608 (2021).
\bibitem{AngeliSontag2003} D. Angeli and E. D. Sontag, Monotone control systems, \emph{IEEE Trans. Automat. Control} \textbf{48}, 1684--1698 (2003).
\bibitem{Deutsch2004ReturnPoint} J. M. Deutsch, A. Dhar and O. Narayan, Return to return point memory, \emph{Phys. Rev. Lett.} \textbf{92}, 227203 (2004).
\bibitem{Schlogl1972} F. Schl\"ogl, Chemical reaction models for non-equilibrium phase transitions, \emph{Z. Phys.} \textbf{253}, 147--161 (1972).
\bibitem{Tero2007Physarum} A. Tero, R. Kobayashi and T. Nakagaki, A mathematical model for adaptive transport network in path finding by true slime mold, \emph{J. Theor. Biol.} \textbf{244}, 553--564 (2007).
\bibitem{CodiscoveryCompanion} V. A. Baulin, Convergent derivations across scientific fields, manuscript in preparation (2026).
\bibitem{Cao2015Oscillations} Y. Cao, H. Wang, Q. Ouyang and Y. Tu, The free-energy cost of accurate biochemical oscillations, \emph{Nat. Phys.} \textbf{11}, 772--778 (2015).
\bibitem{BaratoSeifert2016Clocks} A. C. Barato and U. Seifert, Cost and precision of Brownian clocks, \emph{Phys. Rev. X} \textbf{6}, 041053 (2016).
\bibitem{DommersnesFournier1999} P. G. Dommersnes and J.-B. Fournier, N-body study of anisotropic membrane inclusions: membrane mediated interactions and ordered aggregation, \emph{Eur. Phys. J. B} \textbf{12}, 9--12 (1999).
\bibitem{Meeussen2016Geared} A. S. Meeussen, J. Paulose and V. Vitelli, Geared topological metamaterials with tunable mechanical stability, \emph{Phys. Rev. X} \textbf{6}, 041029 (2016).
\bibitem{Landauer1961} R. Landauer, Irreversibility and heat generation in the computing process, \emph{IBM J. Res. Dev.} \textbf{5}, 183--191 (1961).
\bibitem{Frank1951} F. C. Frank, Crystal dislocations.---Elementary concepts and definitions, \emph{Philos. Mag.} \textbf{42}, 809--819 (1951).
\bibitem{KlemanFriedel2008} M. Kleman and J. Friedel, Disclinations, dislocations, and continuous defects: A reappraisal, \emph{Rev. Mod. Phys.} \textbf{80}, 61--115 (2008).
\bibitem{Bilby1955} B. A. Bilby, R. Bullough and E. Smith, Continuous distributions of dislocations: a new application of the methods of non-Riemannian geometry, \emph{Proc. R. Soc. Lond. A} \textbf{231}, 263--273 (1955).
\bibitem{Berry1984} M. V. Berry, Quantal phase factors accompanying adiabatic changes, \emph{Proc. R. Soc. Lond. A} \textbf{392}, 45--57 (1984).
\bibitem{ShapereWilczek1989} A. Shapere and F. Wilczek, Geometry of self-propulsion at low Reynolds number, \emph{J. Fluid Mech.} \textbf{198}, 557--585 (1989).
\bibitem{Kibble1976} T. W. B. Kibble, Topology of cosmic domains and strings, \emph{J. Phys. A} \textbf{9}, 1387--1398 (1976).
\bibitem{Zurek1985} W. H. Zurek, Cosmological experiments in superfluid helium?, \emph{Nature} \textbf{317}, 505--508 (1985).
\bibitem{Chuang1991Cosmology} I. Chuang, R. Durrer, N. Turok and B. Yurke, Cosmology in the laboratory: defect dynamics in liquid crystals, \emph{Science} \textbf{251}, 1336--1342 (1991).
\bibitem{Hahn1950} E. L. Hahn, Spin echoes, \emph{Phys. Rev.} \textbf{80}, 580--594 (1950).
\bibitem{NanorodCompanion} V. A. Baulin, Geometric frustration and orientational memory in a gold-rod membrane model, manuscript in preparation (2026).
\bibitem{FieldBridge} FieldBridge, \url{https://github.com/synthetix-institute/fieldbridge}, commit 7006641, PolyForm Noncommercial License 1.0.0 (2026).
\bibitem{CarrionVazquez1999} M. Carrion-Vazquez \emph{et al.}, Mechanical and chemical unfolding of a single protein: a comparison, \emph{Proc. Natl. Acad. Sci. U.S.A.} \textbf{96}, 3694--3699 (1999).
\bibitem{Borgia2008} A. Borgia, A. Steward and J. Clarke, An effective strategy for the design of proteins with enhanced mechanical stability, \emph{Angew. Chem. Int. Ed.} \textbf{47}, 6900--6903 (2008).
\bibitem{CarrionVazquez2003} M. Carrion-Vazquez, H. Li, H. Lu, P. E. Marszalek, A. F. Oberhauser and J. M. Fernandez, The mechanical stability of ubiquitin is linkage dependent, \emph{Nat. Struct. Biol.} \textbf{10}, 738--743 (2003).
\bibitem{Schlierf2004} M. Schlierf, H. Li and J. M. Fernandez, The unfolding kinetics of ubiquitin captured with single-molecule force-clamp techniques, \emph{Proc. Natl. Acad. Sci. U.S.A.} \textbf{101}, 7299--7304 (2004).
\bibitem{Rippard2011} W. Rippard, R. Heindl, M. Pufall, S. Russek and A. Kos, Thermal relaxation rates of magnetic nanoparticles in the presence of magnetic fields and spin-transfer effects, \emph{Phys. Rev. B} \textbf{84}, 064439 (2011).
\bibitem{Irie2010} M. Irie, Photochromism of diarylethene molecules and crystals, \emph{Proc. Jpn. Acad. Ser. B} \textbf{86}, 472--483 (2010).
\bibitem{ElAmili2012} A. El Amili, G. Gredat, M. Alouini, I. Sagnes and F. Bretenaker, Experimental study of the delayed threshold phenomenon in a class-A VECSEL, \emph{Eur. Phys. J. Appl. Phys.} \textbf{58}, 10501 (2012).
\bibitem{RoquesCarmes2023} C. Roques-Carmes \emph{et al.}, Biasing the quantum vacuum to control macroscopic probability distributions, \emph{Science} \textbf{381}, 205--209 (2023).
\bibitem{Anquez2016} M. Anquez \emph{et al.}, Quantum Kibble--Zurek mechanism in a spin-1 Bose--Einstein condensate, \emph{Phys. Rev. Lett.} \textbf{116}, 155301 (2016).
\bibitem{PotvinTrottier2016} L. Potvin-Trottier, N. D. Lord, G. Vinnicombe and J. Paulsson, Synchronous long-term oscillations in a synthetic gene circuit, \emph{Nature} \textbf{538}, 514--517 (2016).
\bibitem{Kwakernaak2023} L. J. Kwakernaak and M. van Hecke, Counting and sequential information processing in mechanical metamaterials, \emph{Phys. Rev. Lett.} \textbf{130}, 268204 (2023).
\bibitem{Liu2024} J. Liu \emph{et al.}, Controlled pathways and sequential information processing in serially coupled mechanical hysterons, \emph{Proc. Natl. Acad. Sci. U.S.A.} \textbf{121}, e2308414121 (2024).
\bibitem{Mungan2019} M. Mungan, S. Sastry, K. Dahmen and I. Regev, Networks and hierarchies: how amorphous materials learn to remember, \emph{Phys. Rev. Lett.} \textbf{123}, 178002 (2019).
\bibitem{Pierce2005} M. S. Pierce \emph{et al.}, Disorder-induced microscopic magnetic memory, \emph{Phys. Rev. Lett.} \textbf{94}, 017202 (2005).
\end{thebibliography}
\end{document}